**Deciphering Sub-Neptune Atmospheres: New Insights from Geochemical Models of TOI-270 d**

Christopher R. Glein[1*], Xinting Yu (余馨婷)[2], Cindy N. Luu[2]

[1] Space Science Division, Space Sector, Southwest Research Institute, 6220 Culebra Road, San Antonio, TX 78238, United States

[2] Department of Physics and Astronomy, University of Texas at San Antonio, 1 UTSA Circle, San Antonio, TX 78249, United States

[*] Corresponding author. Email: christopher.glein@swri.org

**Abstract**

The nature of sub-Neptunes is one of the hottest topics in exoplanetary science. Temperate sub-Neptunes are of special interest because some could be habitable. Here, we consider whether these planets might instead be rocky worlds with thick, hot atmospheres. Can recent JWST observations of TOI-270 d be understood in terms of such a model? We perform thermochemical equilibrium calculations to infer conditions of quenching of C-H-O-N species. Our results indicate apparent $CO_2$-$CH_4$ equilibrium between ~900 and ~1100 K. The CO abundance should be quenched higher in the atmosphere where the equilibrium $CO/CO_2$ ratio is lower, potentially explaining a lack of CO. $N_2$ is predicted to dominate the nitrogen budget. We confirm that the atmosphere of TOI-270 d is strongly enriched in both C and $O_{gas}$ relative to protosolar H, whereas N is likely to be less enriched or even depleted. We attempt to reproduce these enrichments by modeling the atmosphere as nebular gas that extracted heavy elements from accreted solids. This type of model can explain the C/H and $O_{gas}$/H ratios, but despite supersolar C/N ratios provided by solids, the $NH_3$ abundance will probably be too high unless there is a nitrogen sink in addition to $N_2$. A magma ocean may be implied, and indeed the oxygen fugacity of the deep atmosphere seems sufficiently low to support the sequestration of reduced N in silicate melt. The evaluation presented here demonstrates that exoplanetary geochemistry now approaches a level of sophistication comparable to that achieved within our own solar system.

**1. Introduction**

*1.1. TOI-270 d as a Rosetta Stone for Temperate Sub-Neptunes*

Exoplanets with sizes and densities between those of Earth and Neptune are a mystery. These sub-Neptunes occupy a region of mass-radius space that permits them to have degenerate combinations of rock, water, and gas – the main ingredients of planets (e.g., Rogers & Seager 2010). We do not know what these planets are made of. It is important to figure this out so we can gain new insights into the origin and evolution of planets given that sub-Neptunes are the most common class of planets observed elsewhere (Fressin et al. 2013); yet, there are no known sub-Neptunes in the solar system. Sub-Neptunes that orbit within the canonical habitable zone, where liquid water can exist on their surfaces, are even more intriguing. They could expand our notions of planetary habitability if they are massive ocean worlds with relatively thin atmospheres. Madhusudhan et al. (2021) popularized the idea that some of these worlds might have $H_2$-rich atmospheres, making them Hycean worlds (earlier, Pierrehumbert & Gaidos 2011 laid the theoretical foundation for how the greenhouse effect of an $H_2$-rich atmosphere can maintain a liquid water ocean on exoplanet surfaces). On the other hand, the opposite endmember would be a planet with much less water and a thick atmosphere overlying a hot (possibly molten) rocky interior. Such conditions would have drastically different implications for the formation environments of these planets, and they would probably be unable to support life. But a planet does not need liquid water to be interesting; other possibilities can still enhance our understanding of these mysterious planets.

The James Webb Space Telescope (JWST) is delivering new data that can illuminate the nature of temperate[1] sub-Neptunes. The first JWST observations of a temperate sub-Neptune were made on K2-18 b. The atmospheric composition determined by transmission spectroscopy in the near-infrared ($H_2$/He atmosphere with ~1% abundances of $CH_4$ and $CO_2$ and no detectable $NH_3$ or CO) seemed to suggest that K2-18 b is a Hycean world (Madhusudhan et al. 2023; Glein 2024). However, a reanalysis of those data found no statistically significant evidence for the detection of $CO_2$ (Schmidt et al. 2025). This finding may cast doubt on the occurrence of Hycean conditions. Moreover, updated photochemical modeling (Wogan et al. 2024) underscores the difficulty of producing sufficient methane on a Hycean world (see Cooke & Madhusudhan 2024). It has been emphasized that K2-18 b is too close to its star to support liquid water at its surface due to greenhouse heating, inhibited atmospheric convection, and patchy cloud cover around the substellar point (Charnay et al. 2021; Innes et al. 2023; Leconte et al. 2024). Various alternative models are now being explored to interpret JWST observations in terms of the expected environmental setting (Huang et al. 2024; Luu et al. 2024; Rigby et al. 2024; Shorttle et al. 2024; Wogan et al. 2024; Yang & Hu 2024; Schmidt et al. 2025). A common feature of all these models is that the atmosphere becomes hot not far below where molecules can be detected. Yet, a key difficulty in understanding the chemistry of this hot region is that the number of tracers that have been robustly identified is limited to $CH_4$ on K2-18 b.

In 2023 and 2024, JWST observed TOI-270 d (Benneke et al. 2024), which is another temperate sub-Neptune. This planet was discovered by the Transiting Exoplanet Survey Satellite (TESS) (Günther et al. 2019) and was then observed by the Hubble Space Telescope (HST) (Mikal-Evans et al. 2023). It was initially suggested that TOI-270 d could be a Hycean world (Madhusudhan et al. 2021; Holmberg & Madhusudhan 2024), but this appears to be a very remote possibility as TOI-270 d receives more intense instellation than K2-18 b does, which makes it even harder to avoid high atmospheric temperatures on TOI-270 d (Benneke et al. 2024). Nevertheless, what is most remarkable about the data from TOI-270 d is their unprecedented quality. They provide the most robust constraints we have on the atmospheric composition of a temperate sub-Neptune. $CO_2$, $CH_4$, and most critically, $H_2O$, have all been detected. With such a rich data set, chemical modeling can be performed to link the abundances of these species to poorly understood interior conditions and the deeper nature of the planet.

Here, we follow up on the interpretation of Benneke et al. (2024) that TOI-270 d is a metal-rich, miscible-envelope sub-Neptune (hereafter, MetMiSN). This terminology means that the atmosphere contains abundant non-$H_2$/He volatiles (so-called metals) in addition to a significant abundance of $H_2$/He, and everything may be well mixed. We perform geochemical modeling of this scenario to assess its consistency with the observed chemical composition and to shed light on the origin of the metal enrichment. While other models of hot gas chemistry on TOI-270 d have already been developed (Benneke et al. 2024; Yang & Hu 2024), there remains room for further exploration. In particular, the present paper brings new approaches to chemical modeling of exoplanetary atmospheres that are intuitive, hands-on, and visually oriented (see Section 1.2). We also try to place this exoplanet in the context of giant planets and primitive materials in the solar system (Kane et al. 2021). It is hoped that this perspective can complement more computationally intensive approaches. The overall goal is to quantify the geochemical implications of a metal-rich, miscible-envelope scenario for the present state and origin of TOI-270 d's atmosphere.

---

[1] A temperate planet is one that receives a similar stellar irradiance as Earth.

*1.2. Aspiration, Guiding Philosophy, and Structure of this Paper*

Below, we present modeling frameworks that can be used to link telescopic observations to deeper planetary conditions (i.e., a top-down perspective) and build the bulk planetary composition from primitive materials (i.e., a bottom-up perspective). Ideally, these types of models will eventually converge, providing insights into distributions, states, and sources of materials in a novel type of planet. Our intention here is to take a first step down this road by showing how this problem can be set up and where we stand in the vast realm of possibilities. We believe that the data from TOI-270 d are detailed enough that an attempt at making this connection should now be made, and by presenting it in one coherent picture, a greater appreciation of both perspectives may be gained.

Along the way, we introduce geochemically inspired approaches for addressing top-down and bottom-up problems. What distinguishes our approaches from a great deal of past work on exoplanets is that we consider analogue processes/materials to fill in gaps (see Glein et al. 2024 for a similar philosophy). This assumption removes lower-level free parameters, which simplifies problems, allowing us to focus on higher-level implications. Furthermore, we know that the analogue is "real," although its relevance to an exoplanet could be debated. It should be noted that we are not claiming that our approaches are better than others. Instead, we wish to provide additional tools so that the community can approach data in different ways.

In Section 2, we use the observed abundances of $CO_2$, $CH_4$, and $H_2O$ in an $H_2$-rich atmosphere to constrain conditions of thermochemical quenching. A novel aspect of our approach is that we do this simply by finding where curves cross in pressure-temperature space. We then take advantage of constraints on the relative quench temperatures of redox reactions from an Earth analogue to understand the potential implications for CO and N-bearing species (see Section 3). With a full set of quenched species, we calculate element ratios with respect to H in TOI-270 d's atmosphere. These ratios serve as targets in the latter half of this study, where we assess whether simple mixtures of primitive materials can reproduce the observationally derived ratios (see Section 4). Our discussion is centered around the apparent absence of $NH_3$ in TOI-270 d's atmosphere (Benneke et al. 2024; Holmberg & Madhusudhan 2024). The case of $NH_3$ provides an excellent example of why we need to approach data from both the top-down (How much bulk N is allowed by JWST observations?) and bottom-up (How much bulk N should there be?). Lastly, we conclude this paper with a summary of key findings and recommendations for future work.

## 2. Chemical Constraints on Conditions of Methane Quenching

### *2.1. Modeling Approach*

On planets with thick atmospheres, it can be expected that the speciation of volatiles will be strongly influenced by thermochemical equilibrium in the deeper atmosphere. As warm air is transported upward, species ratios are effectively frozen in (i.e., quenched) due to the inhibited kinetics of reequilibration reactions at decreasing temperatures (e.g., Prinn & Barshay 1977; Visscher & Moses 2011). Thermochemistry also leads to recycling of photochemically produced gases from the upper atmosphere, such as $CO_2$ and $N_2$, back to $CH_4$ and $NH_3$, respectively (Yu et al. 2021).

The level at which quenching occurs depends on the vigor of vertical mixing relative to the rate of reequilibration. Upon cooling, a reaction will be quenched once the chemical timescale ($\tau_{chem}$) exceeds that of mixing ($\tau_{mix}$). An air parcel that experiences sluggish mixing (large $\tau_{mix}$) will need to cool more for reaction rates to slow down sufficiently (increasing $\tau_{chem}$) until the quench level is reached. Among other things, elucidation of apparent conditions of quenching in a thick, hot atmosphere offers an opportunity to peer into the interplay between chemistry and dynamics occurring deeper in the atmosphere. If TOI-270 d is a MetMiSN, then it seems reasonable to assume that chemical equilibrium would be reached at some depth. However, the specific conditions of this equilibrium are not well-understood (see Benneke et al. 2024; Yang & Hu 2024). Here, a key goal is to set constraints on what the observed composition implies for pressure-temperature conditions of quenching.

$CO_2$-$CH_4$ speciation, the two species best constrained by JWST data, can provide a useful constraint on the relationship between the total pressure ($P$) and temperature ($T$) of equilibrium. Strictly speaking, it should be referred to as an apparent equilibrium since this paper does not attempt to demonstrate that inferred conditions are plausible in terms of kinetic-dynamical considerations (Moses 2014). Rather than construct a forward model accounting for reaction and transport rates needed to *predict* the quenched abundances of species (these models are already plentiful; e.g., Moses et al. 2011; Tsai et al. 2017; Yang & Hu 2024), we solve the inverse problem – we use observed abundances to *infer* quench conditions under the assumption that observed abundances reflect a point of last equilibrium. This type of approach is common in studies of solar system objects (e.g., Fegley et al. 1997; Zolotov & Fegley 2000; Glein et al. 2008). The role of photochemistry is discussed in Section 2.2.

Despite being a simplification, the notion of a quenched equilibrium can be a useful starting point (Moses et al. 2013). In a gaseous environment rich in $H_2$, it is convenient to represent chemical equilibrium between $CH_4$ and $CO_2$ by

$$CH_4(g) + 2H_2O(g) \rightleftharpoons CO_2(g) + 4H_2(g), \tag{1}$$

which can be characterized using the mass action equation

$$K_1 \approx \left(\frac{CO_2}{CH_4}\right)\frac{y_{H_2}^4}{y_{H_2O}^2}P^2, \tag{2}$$

where $K_i$ designates the equilibrium constant of the $i$th reaction, and $y_j$ stands for the mole fraction of the $j$th species. Ideal gas behavior is assumed (see Section 3.2). Equation (2) can be rewritten as

$$\log(P) \approx 0.5\log(K_1) + \log(y_{H_2O}) - 0.5\log\left(\frac{CO_2}{CH_4}\right) - 2\log(y_{H_2}), \tag{3}$$

where the standard pressure unit is the bar. We computed the equilibrium constant of Reaction (1) using Gibbs energies from the NIST-JANAF Thermochemical Tables (Chase 1998), and we fit numerical values to the following equation:

$$\log(K_1) = -2.69 - \frac{8190}{T(K)} + 4.142 \times \log(T(K)), \tag{4}$$

which reproduces the data to within ~0.04 log units from 298-1500 K.

Equation (3) allows us to define *P-T* curves of chemical equilibrium consistent with observational data. However, three compositional parameters must be specified. To organize how these parameters are specified, we define low- and high-pressure endmembers (see Table 1). The high-pressure endmember has a maximum $H_2O$ abundance and a minimum $CO_2/CH_4$ ratio. These are the most uncertain parameters whose effects need to be quantified. "Maximum" and "minimum" values are at the ends of the 1σ ranges of Benneke et al. (2024). The low-pressure endmember has two degenerate cases with either the lowest $H_2O$ abundance or the highest $CO_2$ abundance that provide consistency with radiative transfer model constraints on the tropospheric temperature profile, and are compatible with the observational upper limit on the abundance of CO (see below). The low-pressure endmember has the lowest pressure that falls within physical *P-T* profiles described below. The mole fraction of $CH_4$ is kept constant at the mean value of 2% (Benneke et al. 2024) so that only one variable is changed between the high-pressure endmember and each low-pressure case. Nevertheless, changing the mole fraction of $CH_4$ would not affect the independent observational constraint on the $CO_2/CH_4$ ratio (Benneke et al. 2024). Instead, it would only weakly affect $\log(y_{H_2})$ in Equation (3), as $H_2$ is much more abundant than $CH_4$ is in TOI-270 d's atmosphere.

**Table 1.** Major atmospheric composition of TOI-270 d that yields a lower or upper bound on the total pressure of $CO_2$-$CH_4$ equilibrium.

| Compositional Parameter | Low-Pressure Endmember | | High-Pressure Endmember |
|---|---|---|---|
| | Minimum $H_2O$ Abundance | Maximum $CO_2$ Abundance | |
| Mole Fraction of $H_2$ ($y_{H_2}$) | 0.75 [a] | 0.65 [a] | 0.68 [a] |
| Mole Fraction of He ($y_{He}$) | 0.15 [a] | 0.13 [a] | 0.14 [a] |
| Mole Fraction of $H_2O$ ($y_{H_2O}$) | 0.078 [b] | 0.16 [c] | 0.16 [c] |
| Mole Fraction of $CH_4$ ($y_{CH_4}$) | 0.020 [a,c] | 0.020 [a,c] | 0.020 [a,c] |
| Mole Fraction of $CO_2$ ($y_{CO_2}$) | 0.0054 [c] | 0.037 [b] | 0.0054 [c] |
| Mole Fraction of CO ($y_{CO}$) | <0.035 [c] | <0.035 [c] | <0.035 [c] |
| log($CO_2$/$CH_4$ Molar Ratio) | -0.57 [c] | 0.27 [b,d] | -0.57 [c] |

[a] See Section 2.1.
[b] Calculated in Section 2.2.
[c] Benneke et al. (2024).
[d] Slightly less than Benneke et al.'s (2024) 1σ upper limit of 0.33 to ensure consistency with their upper limit on the mole fraction of CO.

While TOI-270 d's atmosphere is thought to be $H_2$-rich (Mikal-Evans et al. 2023), it is difficult to determine the $H_2$ mole fraction from the near-infrared transmission spectrum, as there is a degeneracy between $H_2$ and He (both have similar effects on the mean molecular weight). Nevertheless, the mole fractions of $H_2$ and He can be estimated using the following equation for the atmospheric composition:

$$1 \approx y_{H_2} + y_{He} + y_{H_2O} + y_{CH_4} + y_{CO_2} = (1 + \text{He}/\text{H}_2) y_{H_2} + y_{H_2O} + y_{CH_4} + y_{CO_2}. \quad (5)$$

This equation is an approximation because it is based on the assumption that unobserved species do not constitute a significant fraction of the composition. We can proceed by assuming the He/$H_2$ ratio to be protosolar (0.20; Lodders 2021), as appropriate (to first order) for primordial gas captured from the planet's formation environment (see Bean et al. 2021).

Values of species mole fractions taken from Benneke et al. (2024) or derived by us are given in Table 1. In addition, we eliminate temperatures at which the CO abundance would exceed the detection limit reported by Benneke et al. (2024). A key aspect of doing that is by accounting for the lower quench temperature of CO oxidation (see Appendices A and B). It is well known in terrestrial geochemistry that CO-$CO_2$ equilibrium is easier to reach than is $CO_2$-$CH_4$ equilibrium (e.g., Giggenbach 1987; Seewald et al. 2006). In Appendix A, we show how relative values of reaction quench temperatures can be estimated based on the speciation of gases from fumaroles. It is important to stress that the fumarole analogue only needs to be considered to provide a basis for extrapolating constraints on *P-T* of $CO_2$-$CH_4$ equilibrium to other reaction systems. The analogue does not affect how we approach $CO_2$-$CH_4$ equilibrium. We verify in Appendix A that, in the same environment, the CO-$CO_2$ pair reaches equilibrium after more cooling than does the pair of $CO_2$-$CH_4$. We propose that CO-$CO_2$ equilibrium will be reached at an absolute temperature ~20% lower than that of $CO_2$-$CH_4$ equilibrium, whereas equilibrium between $N_2$ and $NH_3$ will be reached at an absolute temperature ~20% higher than that of $CO_2$-$CH_4$ equilibrium. We show in Appendix B that our assumption of simple scaling of quench temperatures agrees well (to within ~100 K) with the more complex quench model of Zahnle & Marley (2014). These findings introduce opportunities to estimate unknown quench temperatures when only one is known. The CO-$CO_2$ quench temperature can then be used to determine where the CO upper limit from JWST would be exceeded by evaluating CO-$CO_2$ equilibrium at its quench temperature (see Appendix C). Our approach does not require the unknown eddy diffusion coefficient because an empirical analogue is used to approximate relative quench temperatures of different redox reactions, and they are anchored to an observed equilibrium.

To converge on the conditions of equilibrium, we also need *P-T* profiles in the troposphere of TOI-270 d that are physically plausible. We must resort to radiative-convective modeling because existing observations only probe the stratosphere (~0.1-2 mbar; e.g., Rustamkulov et al. 2023). Yang & Hu (2024) recently modeled TOI-270 d and presented a wide range of *P-T* profiles depending on the planet's atmospheric metallicity and bond albedo. To simplify the analysis and allow broad constraints to be obtained, we define cool and warm atmosphere endmembers encompassing Yang & Hu's (2024) results. Note that the cool atmosphere endmember is for Yang & Hu's (2024) 10 times solar metallicity models rather than the extreme case of solar metallicity. The latter is very inconsistent with the observed atmospheric composition of TOI-270 d (Benneke et al. 2024). One should also note that Benneke et al.'s (2024) nominal *P-T* profile is consistent with our warm atmosphere profile. The adopted *P-T* profiles are shown in the next section.

## 2.2. Results and Discussion

From the composition of TOI-270 d's atmosphere and assuming thermochemical equilibrium, pressure-temperature conditions of quenching can be identified. This is shown in Figure 1 where one high- and two low-pressure curves are plotted to define the area of *P-T* consistent with the compositions in Table 1. Two cases of low pressure are presented to estimate limiting values for the mole fractions of $H_2O$ and $CO_2$ farther below. Figure 1 also shows endmembers of *P-T* estimated for TOI-270 d as a MetMiSN (Yang & Hu 2024). The dashed extensions of the $CO_2$-$CH_4$ equilibrium curves indicate equivalent quench temperatures at which the predicted CO mole fraction is too high (see below). *P-T* conditions consistent with available constraints occur in the shaded region between blue-red and black-dark green curves. At quenching of $CO_2$-$CH_4$ equilibrium, pressure is constrained to 0.9-13 bar with temperature ranging from 885 to 1112 K.

**Figure 1.** Inferring *P-T* conditions of transport-induced quenching from limiting cases of apparent equilibrium for $CO_2$-$CH_4$ (low *P* and high *P* curves; see Table 1) and cool (blue curve)/warm (red curve) endmembers based on Yang & Hu's (2024) predicted *P-T* profiles for TOI-270 d. Only tropospheric profiles from the latter study are shown here as the stratosphere is not relevant to thermochemistry. The cyan region where both constraints overlap is the current best estimate for the *P-T* of last equilibrium between $CO_2$ and $CH_4$. Symbols show the estimated temperature range for $CO_2$-$CH_4$ equilibrium (filled symbols), and the range for $N_2$-$NH_3$ equilibrium (open symbols) estimated using our approach for translating quench temperatures (see Appendix A). Each $N_2$-$NH_3$ temperature was calculated by multiplying the corresponding $CO_2$-$CH_4$ temperature (square or circle) by a factor of 1.2. Dashed portions of the equilibrium curves indicate values of $T_{CO_2-CH_4}$ that imply too much CO (Benneke et al. 2024) at the corresponding (lower) temperature of CO-$CO_2$ quenching.

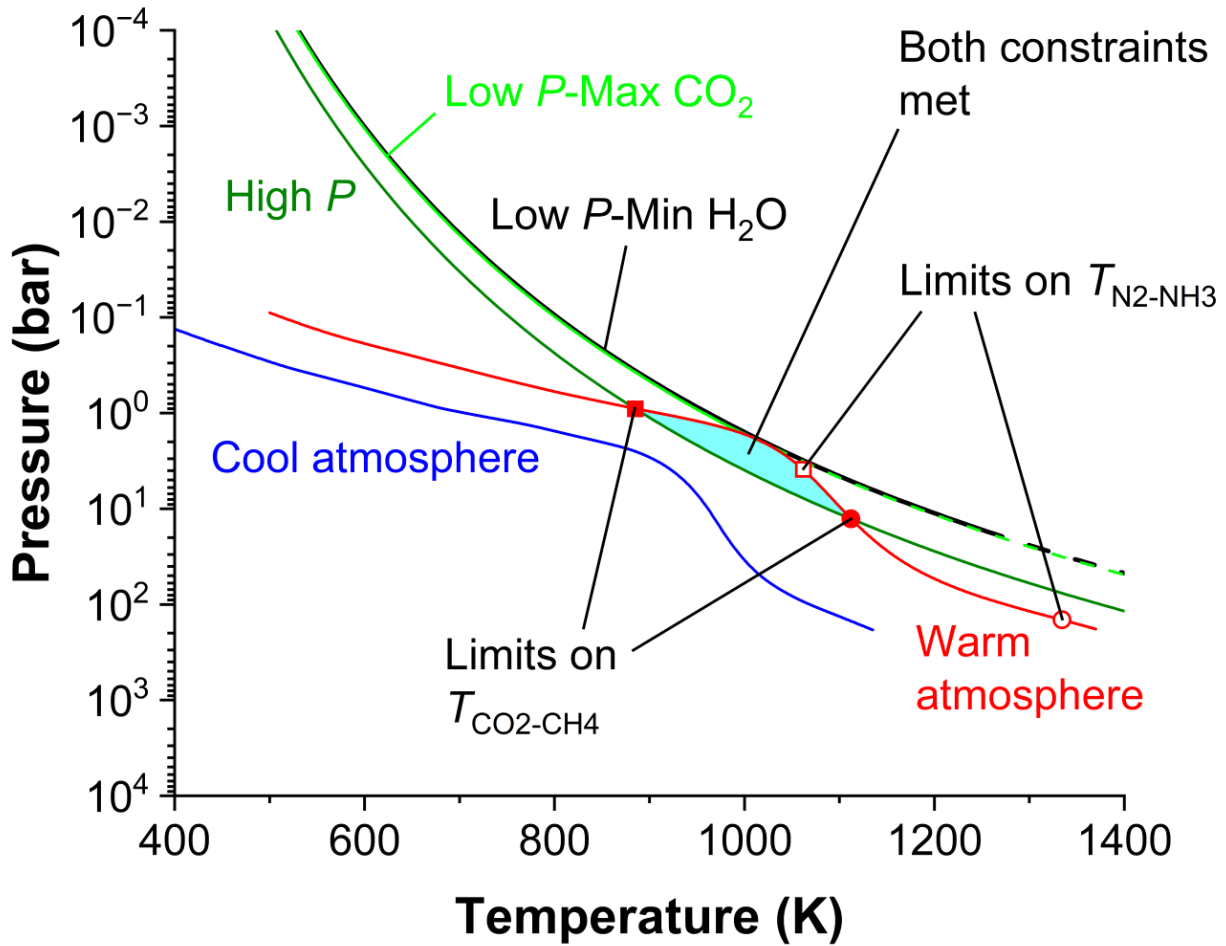

Uncertainty in the magnitude of the planetary heat flow (not modeled here) is unlikely to affect our inferences on the temperature and pressure of $CO_2$-$CH_4$ quenching. A lower planetary heat flow (see Section 3.2) would shift the red and blue curves to the left in Figure 1. But the shift would only occur at pressures higher (deeper) than ~10 bar. Previous radiative-convective modeling showed that the atmospheric temperature profile should not be sensitive to the internal heat flux at altitudes above the ~10 bar pressure level (e.g., Yu et al. 2021; Mukherjee et al. 2024). The quench level of $CO_2$-$CH_4$ equilibrium on TOI-270 d from our analysis (~1-10 bar) generally lies above the level where the internal heat flux begins to matter. We therefore suspect that our upper limit value of 1112 K may be a slight overestimate by perhaps a few tens of kelvins.

The three chemical curves in Figure 1 (note: light green and black overlap) could also be subject to uncertainty. They would be shifted downward if photochemistry produces significant $CO_2$ from $CH_4$ + $H_2O$; then, the original thermochemical value of the $CO_2/CH_4$ ratio should be lower than the observed value. This would allow lower temperatures and pressures of quenching. However, the finding of a high-temperature equilibrium would likely still stand, as a one order-of-magnitude decrease in the $CO_2/CH_4$ ratio would only shift the minimum quench temperature of $CO_2$-$CH_4$ from 885 to 782 K. Moreover, Yang & Hu's (2024) results suggest that TOI-270 d's photospheric[2] composition should be dominated by deeper thermochemistry, and photochemical effects may only become significant at higher altitudes. This situation is the usual expectation for planets around quiet M stars. We note, however, that in the limit of weak vertical mixing, the observable atmosphere would become effectively decoupled from the deep envelope, and no inferences about the interior would be possible from spectroscopy. The dependence of the photospheric composition on $K_{zz}$ (see below) remains to be studied for TOI-270 d.

Inferred temperatures of $CO_2$-$CH_4$ quenching appear to be lower than previously thought. For example, in their nominal model of K2-18 b, Yu et al. (2021) predicted a quench temperature of ~1100 K for methane. This was for an eddy diffusion coefficient $K_{zz} = 10^5$ cm$^2$ s$^{-1}$ in the region where quenching was predicted to occur. Given the similar *P-T* profiles of K2-18 b and TOI-270 d (see Yang & Hu 2024), it is probably appropriate to conclude that the predicted temperature is higher than our derived range (885-1112 K). An implication is that vertical mixing may be slower than what was assumed in Yu et al. (2021), which would provide more time for equilibrium to be reached at lower temperatures. Alternatively, current kinetic models may be missing reactions that produce $CH_4$, underestimating its rate of production. One possibility is that interplanetary dust particles might deliver grains of iron-nickel alloy to the atmosphere. These could catalyze methane production (e.g., Sekine et al. 2006). It should be noted that, on the real planet, the quench temperature will depend on the complex interplay between rates of $CO_2$-$CH_4$ reequilibration and vertical mixing, but from the perspective of our model, it is simply a derived quantity. The critical assumption here is that the observed $CO_2/CH_4$ ratio reflects chemical equilibrium at an idealized single quench point. Nevertheless, it is valuable to know the apparent values of *P* and *T* at this point because they can serve as targets of future efforts to refine kinetic-dynamical forward models for temperate sub-Neptunes.

It should be made clear that the present approach cannot be directly compared to results from Wogan et al. (2024) because our model requires the mole fractions of $CO_2$, $CH_4$, $H_2O$, and $H_2$. The $CO_2/CH_4$ ratio alone is insufficient. Wogan et al. (2024) applied their model to K2-18 b, where water vapor has not been detected by JWST due to cold trapping in the upper atmosphere (Madhusudhan et al. 2023). A restriction on our model for $CO_2$-$CH_4$ speciation is that it cannot be used on "stratified mini-Neptunes" (like K2-18 b; Benneke et al. 2024) since the formation of water clouds on such planets obscures the deep abundance of water. To apply our model, a planet must be warm enough for water vapor to be detected and chemical equilibrium to be reached.

---

[2] The photosphere is the upper layer of an astronomical object (typically, a star) that is transparent to certain wavelengths of light. A transiting planet's photosphere is the region of the atmosphere that produces absorptions that can be observed in the electromagnetic spectrum.

The curve crossings in Figure 1 also have implications for the overall temperature of the ~1-10 bar region of the atmosphere probed by $CH_4$ and $CO_2$. It can be seen that the cool atmosphere endmember is difficult to reconcile with the observed carbon speciation (Figure 1). The temperature profile needs to be shifted to higher temperatures for it to intersect the curves of chemical equilibrium (e.g., Figure E1). The need for a warmer atmosphere is consistent with a high-metallicity atmosphere with an enhanced opacity, and this need also suggests that the bond albedo is low, perhaps close to zero (see Figure B1 in Yang & Hu 2024). We may then deduce that there is a lack of clouds near the substellar point on TOI-270 d. This inference seems to be consistent with the finding of Benneke et al. (2024), based on the terminator, that TOI-270 d's observable atmosphere is largely cloud-free (assuming that we may make a general statement by interpolating between the substellar and terminator zones). Indeed, TOI-270 d's atmosphere is probably always too warm for water clouds to form (unlike on K2-18 b), although other types of clouds could form at unobserved depths.

The present model can be used to explain why JWST did not detect CO on TOI-270 d. The three key factors are the abundances of $CO_2$ and $H_2O$, together with temperature. The abundance of CO will be low when $y_{CO_2}$ is low, $y_{H_2O}$ is high, and $T$ is low. It also helps that the reaction of CO to $CO_2$ is predicted to be quenched at lower temperatures (i.e., 708-890 K) than $CO_2$-$CH_4$ equilibrium (see Appendices A and B). As for compositional relationships, we find that the lack of CO on TOI-270 d does not provide a strong constraint on the water abundance. Rather, Reaction (1) sets a stricter limit on water abundance. If the mole fraction of $H_2O$ were decreased below 7.8%, then the black curve in Figure 1 would shift upward and no longer overlap the red curve. In contrast, the abundance of $CO_2$ must not be too high; otherwise, CO would have been seen by JWST. The light green curve's transition from solid to dashed illustrates this test of chemical consistency. If the mole fraction of $CO_2$ were greater than 3.7%, then the light green curve would become dashed before entering the region bounded by physical *P-T* profiles. Furthermore, the curve would be shifted too far up for cases with more than 4.0% $CO_2$ by volume. A lower limit on $y_{CO}$ can be found at the 885 K intersection of the high-pressure curve and the warm atmosphere profile (Figure 1). The value is 0.26%, or 13 times less than the present detection limit (3.5%; Benneke et al. 2024).

We think that it is most chemically plausible for CO-$CO_2$ to have a lower quench temperature than $CO_2$-$CH_4$ does in an atmosphere that cools with increasing altitude (fewer chemical bonds are broken and formed in the former reaction, making it faster). However, in the limit of very large $K_{zz}$, if one were to assume identical quench temperatures at the lowest quench temperature for $CO_2$-$CH_4$ (885 K), then the predicted mole fraction of CO would be >0.90%, which still agrees with the observational constraint from JWST. These findings demonstrate the sufficiency of equilibrium models of hot gas chemistry to explain the observed carbon speciation on TOI-270 d. Thus, there is no need to invoke Hycean world chemistry to explain a lack of CO detection.

From the distribution of compounds measured by Benneke et al. (2024) and reanalyzed in this work, the elemental ratios C/H and $O_{gas}$/H can be constrained. $O_{gas}$ is oxygen that is not bound to metal cations (Na-K-Mg-Ca-Ti-Al-Si) in silicates (e.g., Fonte et al. 2023). We treat it like a new element. It is important to perform an independent evaluation that carefully considers the potential effects of CO. In Appendix D, we add up all the C, H, and O atoms in $CH_4$, $CO_2$, CO, $H_2$, and $H_2O$, considering a worst-case scenario in which all uncertainties align, and a scenario in which the uncertainty in elemental ratios is driven by the 1σ range of $CO_2$. We settle on the geometric mean of these scenarios as a practical compromise that is sufficient for the purpose of subsequent geochemical modeling. For TOI-270 d, we find a C/H ratio (atom/atom) of 0.016-0.080 (44-220 × protosolar[3]) and $O_{gas}$/H is suggested to be between 0.059 and 0.17 (110-320 × protosolar). These ranges overlap with results (C/H: 0.027-0.080, $O_{gas}$/H: 0.074-0.165) from the quench model of Benneke et al. (2024), but our analysis allows smaller enrichments in both C/H and $O_{gas}$/H.

### 3. Estimating how $N_2$ Production Affects the Upper Limit on the Total Nitrogen Abundance

#### 3.1. Modeling Approach

With constraints on conditions of $CO_2$-$CH_4$ quenching, we can evaluate how corresponding conditions would affect the equilibrium speciation of nitrogen between $NH_3$ and $N_2$. This type of evaluation is essential to constrain the bulk abundance of nitrogen (i.e., the N/H ratio) (Ohno & Fortney 2023a, 2023b). We can now build upon empirical constraints on *P-T* conditions in a temperate sub-Neptune atmosphere to better understand its nitrogen chemistry. Having a deeper understanding of the N/H ratio will give us a clue toward addressing the mystery of missing ammonia in the atmospheres of temperate sub-Neptunes (Madhusudhan et al. 2023; Cabot et al. 2024). Benneke et al. (2024) got us halfway there in deriving an upper limit on the mole fraction of $NH_3$ ($5.4\times10^{-5}$) in TOI-270 d's atmosphere (we use Benneke et al.'s (2024) species abundances rather than those in Holmberg & Madhusudhan (2024), as the former authors analyzed JWST data covering a wider wavelength range). Here, we assume that the photochemically depleted zone for $NH_3$ lies entirely above the planet's photosphere (Yang & Hu 2024) – this allows the upper limit to be applied to the deeper atmosphere (see Section 4.2). Yet, even in this case, JWST cannot easily detect $N_2$, the other stable form of nitrogen in planetary atmospheres. Therefore, the mole fraction of $N_2$ must be estimated indirectly, so that we can then calculate the N/H ratio.

Chemical equilibrium between $NH_3$ and $N_2$ can be represented by

$$2NH_3(g) \rightleftharpoons N_2(g) + 3H_2(g), \tag{6}$$

with an equilibrium constant related to the gas phase composition as shown below:

$$K_6 \approx \frac{y_{N_2} y_{H_2}^3}{y_{NH_3}^2} P^2. \tag{7}$$

Numerical data from NIST-JANAF (Chase 1998) were found to be consistent with

$$\log(K_6) = 2.59 - \frac{4554}{T(\text{K})} + 2.801 \times \log(T(\text{K})). \tag{8}$$

This regression equation reproduces the data to better than ~0.04 log units (298-1500 K). Note that the temperature of $N_2$-$NH_3$ equilibrium in Equation (8) is suggested to be ~1.2 times higher than that where $CO_2$-$CH_4$ equilibrium is quenched, based on what is known about gas quenching in an analogue environment on Earth (see Appendices A and B). The expectation of higher-temperature quenching of $N_2$-$NH_3$ equilibrium is consistent with previous thermochemical kinetics model results, such as those in Zahnle & Marley (2014) and Yu et al. (2021).

---

[3] We adopt the protosolar abundances of Lodders (2021): C/H = $3.6\times10^{-4}$, $O_{gas}$/H = $5.3\times10^{-4}$, and N/H = $8.7\times10^{-5}$. The oxygen abundance is expressed in terms of $O_{gas}$ (Lodders 2004), as oxygen atoms in rock/magma/silicate clouds may not be labile enough to contribute to the inventory of atmospheric gases.

The next step in solving for the nitrogen speciation is to rearrange Equation (7), expressing the mole fraction of $N_2$ in terms of that of $NH_3$:

$$y_{N_2} \approx \frac{K_6}{y_{H_2}^3 P^2} y_{NH_3}^2 = A y_{NH_3}^2, \quad (9)$$

where auxiliary variable $A$ is introduced for a cleaner presentation. A mass balance constraint is also needed. The atomic ratio of N/H is given by

$$\frac{N}{H} \approx \frac{2 y_{N_2} + y_{NH_3}}{2 y_{H_2} + 2 y_{H_2O} + 4 y_{CH_4} + 3 y_{NH_3}} = \frac{2 y_{N_2} + y_{NH_3}}{B + 3 y_{NH_3}}, \quad (10)$$

in which a second auxiliary variable ($B$) is substituted in. $A$ and $B$ are treated as being independent of nitrogen speciation, which holds when the mole fractions of non-nitrogen species are nearly the same in both N-bearing and N-free systems (i.e., the N/H ratio must be small). Equations (9) and (10) can be combined to derive the following quadratic equation:

$$2 A y_{NH_3}^2 + \left(1 - 3 \times \frac{N}{H}\right) y_{NH_3} - B \times \frac{N}{H} \approx 0. \quad (11)$$

We can solve Equation (11) to predict the mole fraction of $NH_3$ along a $P$-$T$ profile. This evaluation also yields the mole fraction of $N_2$. Appendix E details how we obtained relevant values of $A$ and $B$. Briefly, we respeciated bulk C-O-H-He compositions corresponding to the cases in Table 1. Mixtures should be respeciated because temperature and pressure will be higher where $N_2$-$NH_3$ quenching occurs than where the observed $CO_2$-$CH_4$ speciation is set. The mole fraction of $H_2$ decreases slightly in response to these changes. We used the online version of NASA Chemical Equilibrium with Applications (Gordon & McBride 1994; McBride & Gordon 1996) to perform these speciation calculations. Once values of $A$ and $B$ have been calculated, the N/H ratio can be adjusted until it is compatible with the $NH_3$ detection limit on TOI-270 d.

### 3.2. Results and Discussion

$P$-$T$ conditions where chemical equilibrium can be reached dictate the speciation of nitrogen. Quenching of $N_2$-$NH_3$ equilibrium is shifted to higher temperatures and pressures relative to $CO_2$-$CH_4$ equilibrium (see Appendices A and B). As a nominal case, we performed calculations of nitrogen speciation along the warm atmosphere profile from Figure 1; these $P$-$T$ are consistent with the observed carbon speciation of TOI-270 d between the solid symbols in Figure 1. Conditions of apparent $CO_2$-$CH_4$ equilibrium were translated to those corresponding to deeper quenching of $N_2$-$NH_3$ equilibrium (1062-1334 K, 3.9-145 bar; between the open symbols in Figure 1). Equation-of-state calculations using the NIST REFPROP program (Huber et al. 2022) confirmed that gases at these temperatures and pressures are nearly ideal (to within ~4%). We also quantified how the nitrogen speciation would be affected if the temperature profile were uniformly cooler by 50 K. This intermediate temperature profile still falls within the consistency space for the carbon speciation if the atmosphere of TOI-270 d is relatively rich in $H_2O$ and poor in $CO_2$ (Figure E1).

To not exceed Benneke et al.'s (2024) detection limit for $NH_3$ on TOI-270 d, we find that the atomic N/H ratio where quenching occurs should be less than $50\times10^{-4}$ (Figure 2a). This is a conservative upper limit obtained at the lowest quench *P-T*. More total nitrogen can be accommodated if more of it speciates to $N_2$. The formation of $N_2$ is favored at higher temperatures and lower pressures (see Equations (8) and (9)). If conditions are such that $N_2$ production is less than optimal, then the N/H ratio will need to be lower. Hence, a more robust upper limit for the N/H ratio is $\sim2\times10^{-4}$. This is meant to serve as a practical upper limit that permits some variability in the quench temperature for nominal conditions (solid curves in Figure 2a) and in the thermal structure of the atmosphere (Figure E3a). Since the protosolar N/H ratio is $8.7\times10^{-5}$ (Lodders 2021), the conservative upper limit on N/H in TOI-270 d's troposphere is 60 × protosolar; the practical upper limit is ~2 × protosolar. These values are larger than what we would derive from the abundance of $NH_3$ alone because our model predicts that there should be more $N_2$ than $NH_3$ in the quenched composition of TOI-270 d (Figures 2b and E3b). Yet, the abundance of $N_2$ is predicted to be below 0.4% by volume and may even be <0.01%. This is indeed a nitrogen-poor atmosphere.

**Figure 2.** (a) Abundance of $NH_3$ and (b) $N_2/NH_3$ ratio from our model of differential quenching of $N_2$-$NH_3$ and $CO_2$-$CH_4$ equilibria on TOI-270 d. These predictions are for conditions along the warm atmosphere profile between the open symbols in Figure 1 (see also Figure E3). Colored curves show effects of varying the C-O-H composition of the atmosphere (see Table 1). Dashed and solid curves in (a) can be used to set conservative and practical upper limits, respectively, on the N/H ratio based on the non-detection of $NH_3$ by JWST (Benneke et al. 2024).

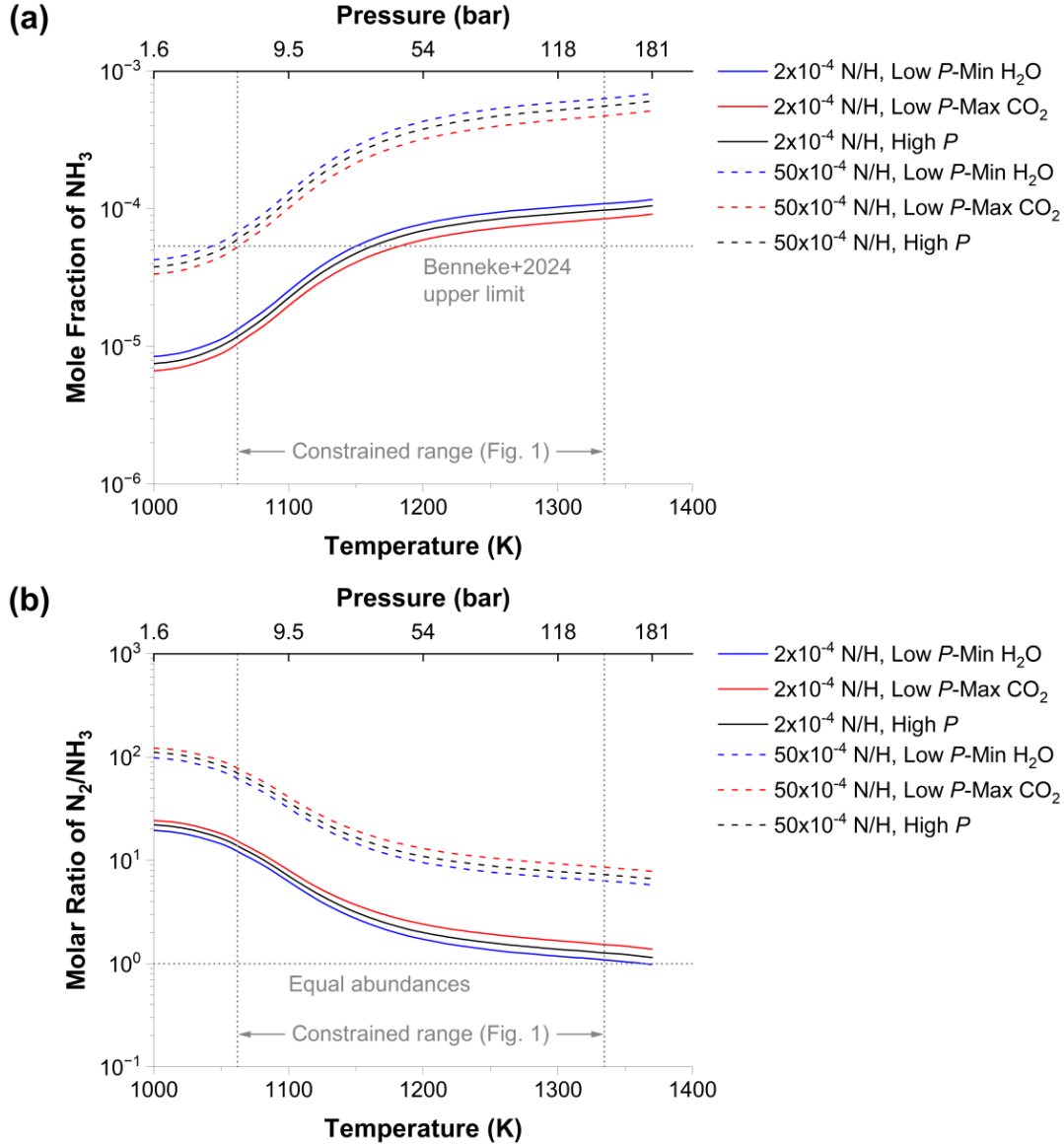

Our predictions of the nitrogen speciation in TOI-270 d's atmosphere should be seen as approximations that accomplish three goals. First, they demonstrate how a new method can be used to extrapolate measurements of the carbon speciation to another element that has yet to be observed. Second, they emphasize the potentially important role of $N_2$ as a hidden reservoir of nitrogen atoms in the thermochemically active region of TOI-270 d's atmosphere. Third, they provide first-order estimates of the bulk N content of the atmosphere of TOI-270 d, which is needed to begin exploring how the present atmosphere could relate to different formation scenarios (see Section 4). In detail, however, our predictions come from a model that is affected by uncertainties in the quench temperature of $N_2$-$NH_3$ equilibrium and temperatures in the deeper atmosphere. A full evaluation of these uncertainties is beyond the scope of this work that is focused on big-picture ideas and modeling frameworks, but below we discuss key issues and trends.

The first issue is potential inaccuracy in the temperature of $N_2$-$NH_3$ quenching. In terms of physical processes, vertical mixing has a crucial effect on quench temperatures of different reactions because different reactions proceed at different rates. As an example, rapid vertical mixing (small $\tau_{mix}$) implies quenching at high temperatures so that the timescale of reequilibration is also short (small $\tau_{chem}$). In our model, the effect of vertical mixing is "baked in" by determining the apparent quench temperature for $CO_2$-$CH_4$ (Figure 1). A high temperature would imply rapid upward transport, even though no transport parameter (e.g., $K_{zz}$) is explicitly calculated. We then estimate the temperature of $N_2$-$NH_3$ quenching by applying an empirical scaling factor. This simplification seems to work surprising well (e.g., Figure B1). It should be noted, however, that at $T_{CO_2-CH_4}$ between 885 and 1112 K (i.e., the inferred range on TOI-270 d) our model may overestimate $T_{N_2-NH_3}$ by up to ~100 K (Figure B1b). On the other hand, the kinetics of $N_2$-$NH_3$ reequilibration are very slow (e.g., Tsai et al. 2018). For the sake of argument, if we assume that our model overestimates $T_{N_2-NH_3}$, then the plateau on the left side of Figure 2b may be most relevant. The equilibrium abundance of $N_2$ would increase somewhat but not by enough to account by itself for "missing ammonia" discussed in Section 4.2.

Another source of uncertainty is the temperature profile below the ~10 bar level. TOI-270 d's atmosphere could be richer in total N if the deeper atmosphere is hotter than what was considered. We used a large set of *P-T* profiles for TOI-270 d from Yang & Hu (2024) to define our bounds. Their calculations were based on an internal heat flux corresponding to an intrinsic temperature ($T_{int}$) of 60 K (i.e., similar to Neptune's $T_{int}$ of ~53 K). Note that we are not suggesting that the interior of TOI-270 d is compositionally or geodynamically similar to that of Neptune. Instead, we are merely pointing out that the previously published thermal structure of its atmosphere was based on an overall planetary heat flux that is similar to Neptune's, although the sources of heat and transport mechanisms may differ. If $T_{int}$ were higher than what was assumed, then $N_2$ formation would be more favorable at depths where quenching occurs, allowing more total N to be present (Fortney et al. 2020). However, this seems unlikely as TOI-270 d is less massive than Neptune, and TOI-270 d may not experience substantial tidal heating. It is the outermost known planet in its system and has a low eccentricity (Kaye et al. 2022). A lower $T_{int}$ more similar to Earth's (~35 K; Lucazeau 2019) is more likely, which would imply that even our practical upper limit on N/H may be somewhat conservative. One must venture deeper into the atmosphere to reach the quench temperature with decreasing $T_{int}$ because of a lower internal heat flux. However, higher *P* favors a higher abundance of $NH_3$, which weakens $N_2$ as a nitrogen sink, thus reinforcing the finding that TOI-270 d has a nitrogen-poor atmosphere.

With constraints on the enrichments of C, $O_{gas}$, and N, we can examine how TOI-270 d's atmosphere compares to the atmospheres of Jupiter, Saturn, Uranus, and Neptune. Table F1 provides the data. In making these comparisons, we must keep in mind that the internal structure of TOI-270 d (mostly rock by mass) is much different from those of the solar system's giant planets. Nevertheless, this difference does not preclude a similar origin of their atmospheres. Within the atmosphere, TOI-270 d is thought to have a substantial $H_2$ inventory (65-75 mol %; Table 1). The most likely source of this inventory is primordial gas that was captured from the planet's formation environment. Our solar system's giant planets also captured nebular gas, although they captured more than TOI-270 d did given that they are more massive and have lower bulk metallicities (Guillot et al. 2023). In addition, it may be of interest to contextualize results for TOI-270 d by showing them on plots with the giant planets that we know best. We see this as an opportunity to spur further discussion.

Graphical comparisons of how elemental enrichments vary with the mass of the planet are shown in Figure 3. These enrichments apply to atmospheres and do not necessarily apply to deep interiors (see Guillot et al. 2023). Two patterns emerge in Figure 3. First, TOI-270 d's C and $O_{gas}$ enrichments are consistent with trends seen in the solar system, where the enrichment decreases with increasing planet mass (Kreidberg et al. 2014). The C enrichment shows closer agreement, as we might expect since the solar system trend line for C is more reliably based on observational data for all four planets (see Table F1). $O_{gas}$ enrichments on our giant planets are mostly estimated via tracers of water (CO, D/H ratio) because their upper atmospheres are too cold to show the bulk abundance of water (Weidenschilling & Lewis 1973). At Jupiter, Juno has measured deep water, but so far, only in the equatorial zone (Li et al. 2024). The broad agreement of TOI-270 d's C and $O_{gas}$ enrichments with gas giant trend lines from the solar system suggests that at least some volatile elements in TOI-270 d's atmosphere share a similar origin as those found in the atmospheres of Jupiter, Saturn, Uranus, and Neptune. This seems like the simplest explanation, although it follows from an empirical extrapolation.

**Figure 3.** C-O$_{gas}$-N enrichments vs. mass in TOI-270 d's atmosphere compared to the atmospheres of solar system giant planets. Values were derived from molecular mixing ratios and normalized to the protosolar abundances of Lodders (2021). Thus, they are internally consistent. The mass of TOI-270 d was taken from Van Eylen et al. (2021). Colored regions represent the area bounded by the standard error of the fit. Our intent here is to place TOI-270 d in the context of the giant planets that we know best. It is not to make a comprehensive comparison with a larger data set of exoplanets. The latter shows more scatter (see Swain et al. 2024). Note that the ordinate axis in (c) is shifted down by a factor of ten, accentuating the limited N enrichments.

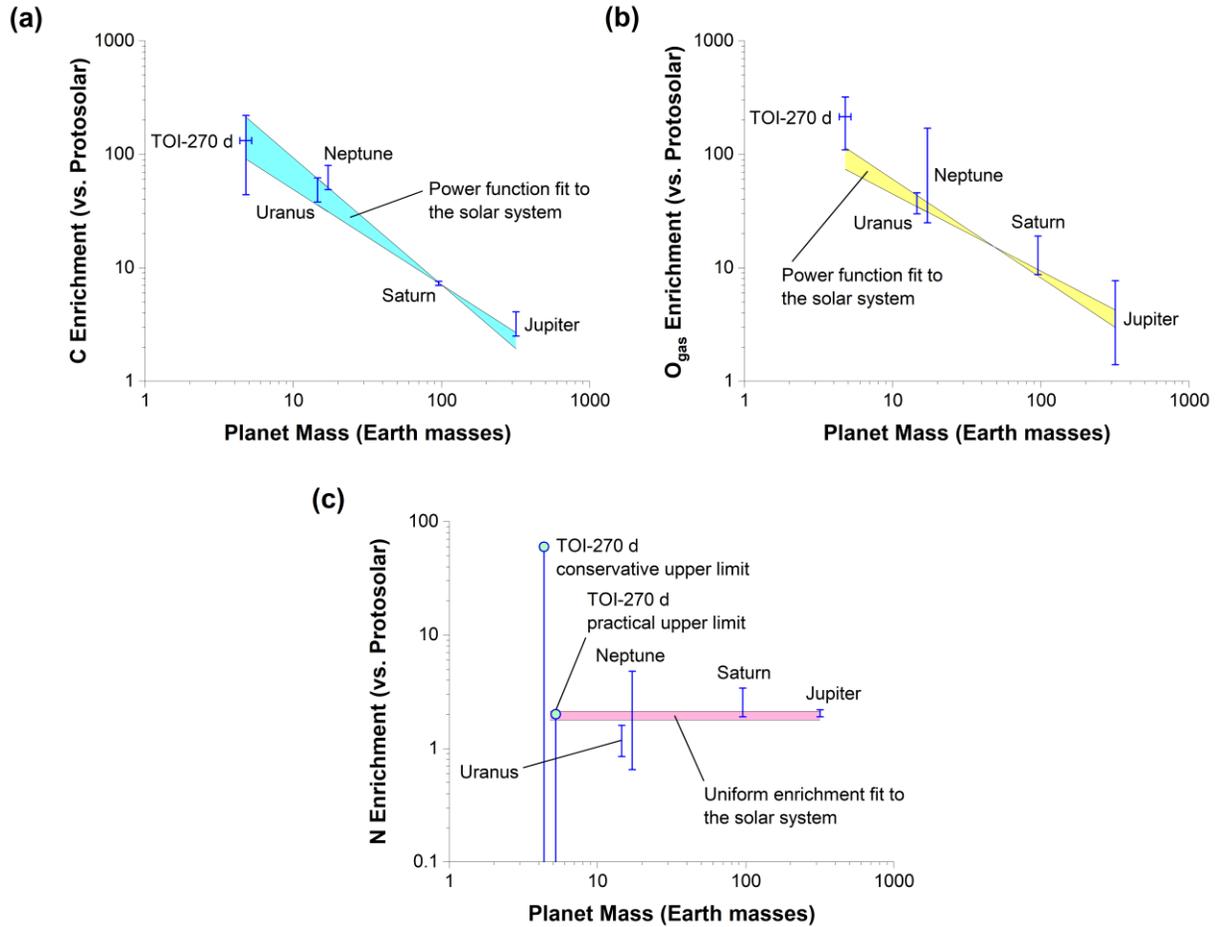

The nitrogen enrichment shows a much different trend (Figure 3c). The simplest fit to the data is a constant enrichment (~2 × protosolar) independent of planet mass. However, this fit should be interpreted with caution because we cannot be sure that available measurements of NH$_3$ on Saturn, Uranus, and Neptune were deep enough to reflect bulk abundances (e.g., Atreya et al. 2019). Nevertheless, there is a hint that N benefits less from a decrease in planetary mass than C and O$_{gas}$ do. Also, the presence of H$_2$S clouds on Uranus and Neptune indicates that their atmospheres have N/S ratios less than unity (Irwin et al. 2018, 2019), whereas the protosolar value is ~5 (Lodders 2021). Therefore, it should not be surprising for an analogous exoplanetary atmosphere to be missing nitrogen, even if the cause is unclear (see Section 4.2). Indeed, our practical upper limit on TOI-270 d's N enrichment is consistent with the constant trend line in Figure 3c. NH$_3$ depletion in sub-Neptune atmospheres may not be used to rule out the existence of a deep atmosphere.

## 4. Toward Understanding the Abundance Pattern of Volatile Elements

### *4.1. Modeling Approach*

Now that we have constrained the enrichments of C, $O_{gas}$, and N in TOI-270 d's atmosphere, the next question is where did these elements come from? While we do not have all the answers, here we would like to introduce a modeling framework that may open up a new direction by beginning to bridge the communities of exoplanetary science and cosmochemistry. With detailed compositional data from solid-dominated planets becoming available, we can benefit from a wealth of information on primitive materials in the solar system as planetary building blocks. The reader should be aware that this section is more speculative than the preceding sections because of lingering uncertainties in the atmospheric composition of TOI-270 d and its relationship to the planet's bulk composition. Nevertheless, exploration can start with informed speculation. There is value in attempting to place atmospheric observations in a cosmochemical framework to identify limitations in the data as well as in our ability to quantify the evolution of volatiles on planets. The approach can also help to guide future interpretations as the quality of observational data improves.

We construct two types of models to explore the origin of TOI-270 d's heavy element enrichments. First, we test whether the conventional model of uniform metallicity enrichment (e.g., Hu 2021) is consistent with elemental abundances in the atmosphere of TOI-270 d. We assume that all elements heavier than He are enriched relative to H by the same factor of *E* times their protosolar ratios from Lodders (2021). We treat carbon as a reference element, as it is the most accessible heavy element on giant planets (oxygen in water can be sequestered in clouds). The enrichment factor of heavy element Z is related to that of C by

$$E_Z = E_C . \qquad (12)$$

A uniform heavy element enrichment could be produced if (1) the planet captured nebular gas that was already enriched in heavy elements (Guillot & Hueso 2006), (2) cold solids with protosolar Z/C ratios contributed to the formation of the planet's atmosphere (Owen et al. 1999), or (3) atmospheric escape removed H but elements heavier than He were retained (Malsky et al. 2023). Furthermore, a combination of processes is not to be excluded.

We also explore an alternative cosmochemical model in which the atmosphere of TOI-270 d was sourced from mixing of protostellar gas with accreted solids having non-solar Z/C ratios (Schaefer & Fegley 2010; Thompson et al. 2021). This approach is agnostic to how volatiles from the second source reach the surface. They may enter the atmosphere via outgassing from the interior (Allègre et al. 1987) and/or be delivered directly by volatile-bearing planetesimals (Müller & Helled 2024). We think that atmospheres in the solar system originated from mixtures (e.g., Miller et al. 2019).

For a two-component model, the following equations can be used to predict the relative abundances of C and Z in a planet's atmosphere:

$$\left(\frac{C}{H}\right)_{atm} = \frac{\mu_H}{\mu_C} \frac{w_{C,gas} f_{gas} + w_{C,solids} f_{solids}}{w_{H,gas} f_{gas} + w_{H,solids} f_{solids}}, \qquad (13)$$

and

$$\left(\frac{Z}{H}\right)_{\text{atm}} = \frac{\mu_H}{\mu_Z} \frac{w_{Z,\text{gas}} f_{\text{gas}} + w_{Z,\text{solids}} f_{\text{solids}}}{w_{H,\text{gas}} f_{\text{gas}} + w_{H,\text{solids}} f_{\text{solids}}}, \tag{14}$$

where $w_{i,j}$ refers to the mass fraction of element $i$ in the $j$th source material, $\mu_i$ corresponds to the molecular weight of element $i$, and $f_j$ stands for the mass fraction of the planet's atmosphere that was contributed by source material $j$. It should be noted that this model considers the relative contributions of accreted gas and solids to be an embedded free parameter, and the model does not require the full volatile content of the planet to be in the atmosphere (e.g., Luo et al. 2024). The two $f$ values in Equations (13) and (14) sum to unity:

$$f_{\text{gas}} + f_{\text{solids}} = 1. \tag{15}$$

By combining Equations (13)-(15), we can study how the enrichment factors of C and Z are related through mixing, given

$$E_Z = \frac{\left(\frac{Z}{H}\right)_{\text{atm}}}{\left(\frac{Z}{H}\right)_{\odot}} \tag{16}$$

(subscript ☉ denotes the protosolar ratio). This can be done by varying $E_C$ and solving for $f_{\text{gas}}$ and $f_{\text{solids}}$ using Equations (13) and (15). We then input the latter values into Equation (14) to predict the $O_{\text{gas}}$ or N enrichment for the mixture of interest.

      Table G1 provides mass fractions in the C-N-O-H-He system for potential analogues of the building blocks of TOI-270 d's atmosphere. We consider a wide range of compositions, which include nebular gas, several types of chondritic meteorites (e.g., Elkins-Tanton & Seager 2008), and a model of comet 67P/Pluto. Our model of comet 67P/Pluto combines spacecraft data on volatile elements from both bodies to derive a bulk composition that is a best estimate for distant icy material in the solar system (see Appendix H). The compositions considered here are likely to be representative across a range of formation environments, including outside the $H_2O$ snow line (Bitsch et al. 2021; Hopp et al. 2022). Our approach is different from the usual approach of developing detailed theoretical astrochemical models of protoplanetary disk chemistry (Öberg et al. 2011; Madhusudhan et al. 2014; Mordasini et al. 2016; Mollière et al. 2022). Instead, we aim to complement those models by focusing on empirical examples of planetary building blocks, as commonly done for Earth (e.g., Marty 2012; Dauphas 2017). In our model, it is assumed that metals that are difficult to reduce will retain oxygen and the remaining oxygen will become $O_{\text{gas}}$. The latter includes O originally bound to Fe, which may produce endogenous water under a thick, $H_2$-rich atmosphere (Kite & Schaefer 2021). Thus, the present approach accounts for and consolidates diverse sources of water. However, our mixing model assumes that the atmosphere faithfully inherits ratios of volatile elements from its building blocks. This should be seen as a simplified endmember rather than as a constraint on fractionation processes, such as preferential removal of certain elements from the atmosphere into silicate melt or metallic phases in the interior (see Suer et al. 2023). As we show in the next section, there is still value in exploring the simplest scenario of direct inheritance as the existence of glaring discrepancies can help us to understand whether observational data require more complex scenarios.

*4.2. Results and Discussion*

Different models for enriching heavy elements can be tested using Figure 4. We see solutions that allow all of our models (not to mention higher-order mixtures) to achieve consistency with inferred ranges of C/H and $O_{gas}$/H on TOI-270 d (Figure 4a). Therefore, we must take care in making specific interpretations. Chondritic material mixed with nebular gas can explain the $O_{gas}$ enrichment where the C enrichment is relatively low (~44-115 × protosolar), while comet 67P/Pluto material or scenarios that maintain a protosolar C/$O_{gas}$ ratio can agree with the data at higher C enrichments (~105-220 × protosolar). The latter include solar composition icy planetesimals (SCIPs) – very cold material that is hypothesized to trap heavy elements except for neon in protosolar proportions (Owen & Encrenaz 2006). Although TOI-270 d may be a rock-rich planet that formed inside the snow line (Benneke et al. 2024; Yang & Hu 2024), we cannot rule out the possibility that it accreted some ice-rich material, perhaps from a limited number of impactors that originated farther out in its protoplanetary disk. Because the atmosphere's mass is relatively small (Benneke et al. 2024), its heavy element inventory can be easily enriched (Fortney et al. 2013).

**Figure 4.** Enrichments in the atomic ratios of (a) $O_{gas}$/H and (b) N/H as functions of C/H enrichment from our model of mixing between primordial gas and various chondrite types or comet 67P/Pluto material (colored curves). These solids may contribute to the origin of planetary atmospheres. Enrichments are with respect to the protosolar abundances of Lodders (2021). More solid material is mixed with $H_2$-He-rich nebular gas from left to right. The atmosphere's C/$O_{gas}$ and C/N ratios eventually converge to slopes defined by accreted solids. Dashed lines highlight where C/$O_{gas}$ and C/N ratios are always fixed at their protosolar values. Shaded boxes show present constraints on volatile element ratios in TOI-270 d's atmosphere along with the atmospheres of giant planets in the solar system (see Table F1). Note the one decade shift in the vertical axis scales, and that a few chondrites are not shown (but indicated) to minimize congestion.

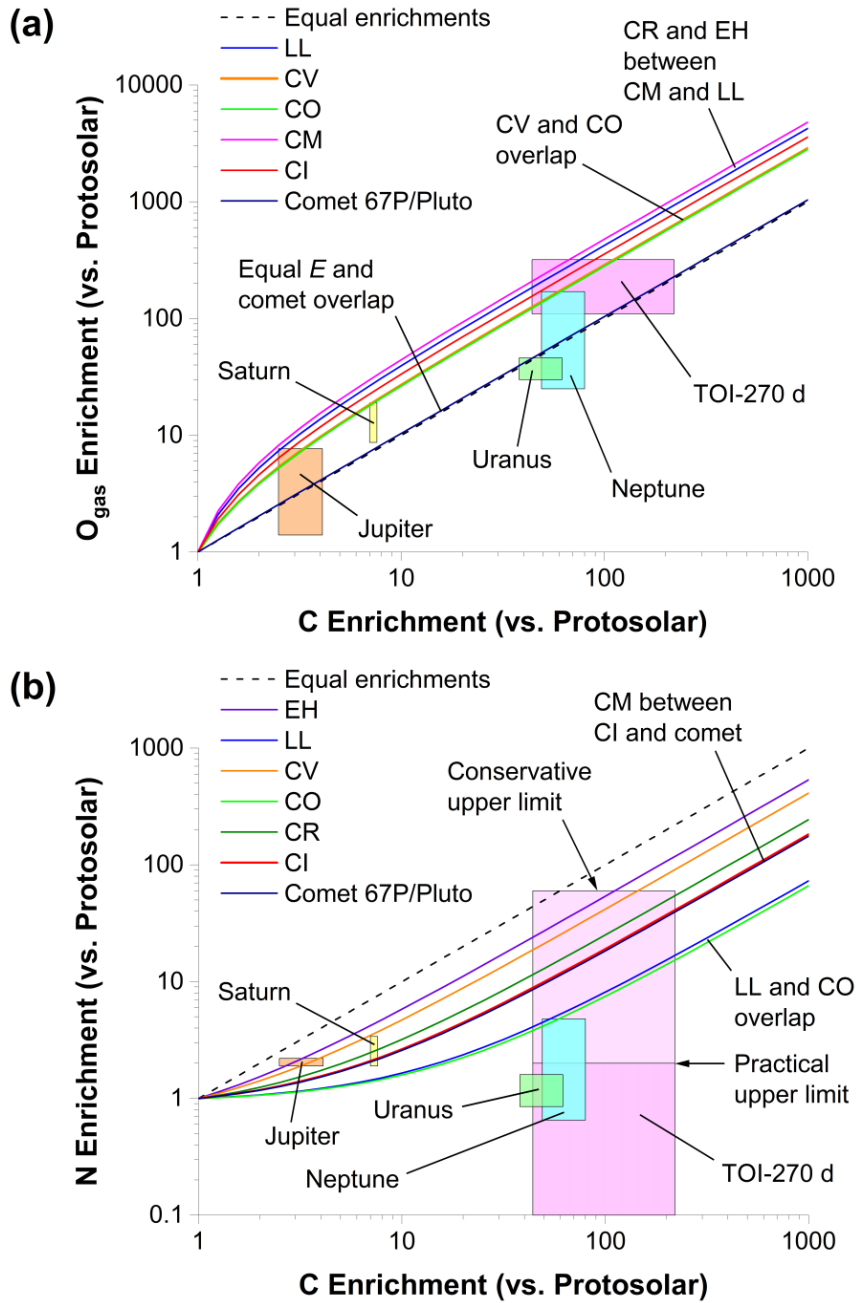

We can also consider the opposite perspective where TOI-270 d's C and $O_{gas}$ enrichments allow the source of heavy elements in its atmosphere to be nominally dry rocks, such as EH and LL chondrites (Figure 4a). Perhaps these elements came from outgassing of interior rocks that were accreted locally in the inner region of the host star's protoplanetary disk. This possibility appears to differ from our giant planets, which tend to cluster near the bottom lines. The latter observation suggests that carbon-rich outer solar system materials were delivered to their atmospheres (Malamud et al. 2024; Mousis et al. 2024). If there is indeed a difference, it may arise simply because of different planet locations (i.e., inner vs. outer disk). In terms of the core accretion paradigm (Pollack et al. 1996), we might assume that TOI-270 d formed as a mostly silicate-metal core that captured a small amount of nebular gas (see Bean et al. 2021), whereas our giant planets apparently accreted cores with more organic matter and/or CO ice (in addition to water ice). Figure 4a emphasizes the importance of more tightly constraining the C enrichment to narrow down the possibilities for the types of materials that supplied volatiles to TOI-270 d. This figure may also be of use to interpret observations of other exoplanets and future data from the solar system's giant planets.

We next evaluate how much of a nitrogen deficit TOI-270 d has. The idea is to use models of possible planetary building blocks to predict how much nitrogen there should be, and then compare these predictions to the upper limits on N/H that we estimated in Section 3.2. The curves in Figure 4b show that the N enrichment should correlate with the C enrichment since solid building blocks contain both elements. It can be seen that all of the model curves cross through the light pink region, indicating that they are consistent with the conservative upper limit on N enrichment. If this upper limit is most relevant, then there does not appear to be a nitrogen deficit. The non-detection of $NH_3$ on TOI-270 d (Benneke et al. 2024; Holmberg & Madhusudhan 2024) can be explained by the combination of variations in the intrinsic inventory of N and $N_2$ production by thermochemistry. Other processes could also be operating (see below), but they are not required. Alternatively, the practical upper limit on N enrichment (dark pink region in Figure 4b) leads to an entirely different interpretation – there is always a nitrogen deficit. This is our preferred interpretation because it seems more likely (see Section 3.2). In this case, TOI-270 d would not be a metal-rich, miscible-envelope sub-Neptune (Holmberg & Madhusudhan 2024), or some additional process is removing nitrogen.

To aid assessments of the plausibility of the latter hypothesis, we derive lower limits on the depletion factor of N, defined as

$$D_N = \frac{(N/H)_{bulk}}{(N/H)_{obs}}, \tag{17}$$

where the bulk ratio comes from one of our cosmochemical models, and the observed ratio is the conservative or practical upper limit. Derived values are given in Table I1. These were calculated over ranges of C/H that provide mutual consistency for all four elements with respect to observational constraints. We find that the equal enrichments model is most discrepant, which may lead to it being disfavored unless there is a potent N-depleting process on TOI-270 d. In contrast, LL and CO chondrite models show only minor discrepancies; these materials seem like promising sources of heavy elements in TOI-270 d's atmosphere.

For TOI-270 d to be a MetMiSN, there likely needs to be an extra process that removes $NH_3$ from its atmosphere (Benneke et al. 2024). Two such processes were proposed initially for K2-18 b, but they also have relevance to TOI-270 d. The first suggestion is that ammonia is missing because it is depleted by photochemistry (Wogan et al. 2024). It is well known that $NH_3$ is sensitive to ultraviolet light (e.g., Kasting 1982). However, the situation may be trickier on temperate planets around M stars. Hu (2021) showed that weak UV output from M stars can allow significant amounts of $NH_3$ to persist at higher altitudes in planetary atmospheres. Indeed, other pre-JWST chemical models of K2-18 b predicted that $NH_3$ should be detectable in the planet's photosphere if it has a deep atmosphere (Tsai et al. 2021; Yu et al. 2021). The non-detection of $NH_3$ on K2-18 b (Madhusudhan et al. 2023) and TOI-270 d (Benneke et al. 2024) therefore raised a red flag.

Wogan et al. (2024) presented a model that could explain this discrepancy. Their model invokes a drastic decrease in the eddy diffusion coefficient $K_{zz}$ (down to $10^3$ cm$^2$ s$^{-1}$) at 1 bar; i.e., between the quench level and photosphere. This leads to much weaker delivery of $NH_3$ from the deep atmosphere, allowing photolysis to deplete $NH_3$ to lower altitudes, potentially extending through the planet's photosphere. However, the photochemical destruction of $NH_3$ should be slow (Tsai et al. 2021) because the host star of TOI-270 d is a quiet M star (Günther et al. 2019). A recent investigation of the parameter space feeding into disequilibrium chemistry underscores how sensitive the photospheric abundance of $NH_3$ is to $K_{zz}$ (Mukherjee et al. 2024). In 1D models, $K_{zz}$ is effectively an empirical parameter (see Komacek et al. 2019). Therefore, it is plausible for $K_{zz}$ to assume values that allow a model to be consistent with the lack of $NH_3$ detection. But what would make the explanation more compelling is if an independent observable can be found that also calls for weak vertical mixing. $SO_2$ may be one such species; it was potentially seen on TOI-270 d and is thought to have a photochemical origin (Benneke et al. 2024). For photochemical gases to have a greater representation in the observed composition, they should be subject to less recycling and slower transport into the interior. Another possible indicator of a strong photochemical influence would be the presence of $C_2H_6$ (Holmberg & Madhusudhan 2024).

A second explanation that has been put forward to explain the lack of $NH_3$ in temperate sub-Neptune atmospheres is that these planets could have oceans of silicate magma at the base of thick atmospheres, and the ocean dissolves nitrogen in a chemically reduced form (Shorttle et al. 2024). A key geochemical parameter that controls the ability of magma to sequester nitrogen is the oxidation state of the magma (e.g., Bernadou et al. 2021). Assuming that magma is present and equilibrates with the lower atmosphere, TOI-270 d presents us with an excellent opportunity to constrain oxidation states on a sub-Neptune since major O-bearing species were measured in its atmosphere (Benneke et al. 2024). What is most important is the $H_2O/H_2$ ratio in the lower atmosphere, which can be related to oxygen fugacity ($f_{O_2}$) as a measure of oxidation state (e.g., Kite et al. 2020). As described in Appendix J, we can use the bulk $O_{gas}/H$ ratio, together with estimates of temperature, pressure, and non-ideal gas behavior, to set broad constraints on $f_{O_2}$. Because oxygen fugacity varies significantly only on a logarithmic scale, even uncertainties that may seem large do not matter much for understanding the redox speciation of nitrogen and other elements.

Figure 5 shows upper and lower limit curves for the oxidation state at the surface of TOI-270 d. These plots also show reference curves for redox conditions described by the iron-wüstite (IW) buffer and states more reduced than IW. It is common in petrology to refer to oxygen fugacities relative to well-characterized states like IW rather than absolute values of $f_{O_2}$. Values of $f_{O_2}$ for IW were calculated using the analytical representation of Hirschmann (2021). It can be deduced that at the temperatures (4000-5000 K) proposed by Benneke et al. (2024) for a magma ocean on TOI-270 d, inferred oxygen fugacities are more reduced than IW and can be as reduced as IW − 4 (i.e., 4 log $f_{O_2}$ units below IW).

**Figure 5.** Oxidation state of a magma ocean at the surface of TOI-270 d at (a) 290 kbar (29 GPa) and (b) 340 kbar (34 GPa). Solid blue curves show upper and lower bounds on oxygen fugacity based on corresponding limits for the ratios of $O_{gas}$/H and $H_2O$/$H_2$ fugacity coefficients (see Appendix J). Dashed blue curves show how these results would change if the ratio of fugacity coefficients were assumed to be unity. $O_2$ itself should not be abundant enough to upset the $O_{gas}$ mass balance because $f_{O_2}$ is much smaller than the total pressure. Red curves indicate $O_2$ fugacities relative to those of the iron-wüstite (IW) buffer. The solid portion of each red curve is where the model of IW was calibrated (Hirschmann 2021), while the dotted portion represents an extrapolation (which exhibits reasonable, quasi-linear behavior with $T^{-1}$). Benneke et al.'s (2024) suggested temperature range for a surface magma ocean is also indicated. Yellow regions highlight geochemical conditions compatible with our analysis while allowing sufficient nitrogen to be removed from the atmosphere, completing a series of processes that may resolve the mystery of missing ammonia.

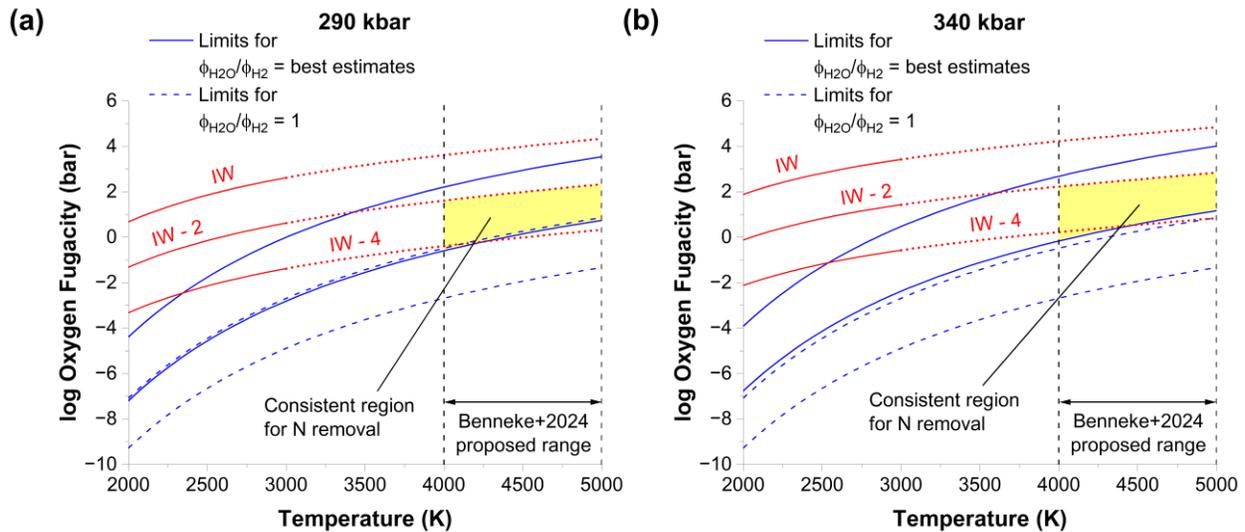

A rule of thumb is that log $f_{O_2}$ should be below ~IW − 2 to decrease the atmospheric abundance of total N by more than an order of magnitude (Rigby et al. 2024; Shorttle et al. 2024). This is roughly how much N sequestration is necessary to reconcile top-down and bottom-up estimates of the nitrogen inventory (Table I1). The lower atmosphere of TOI-270 d can be reduced enough to satisfy this requirement, as indicated by the yellow regions in Figure 5. This finding is independent of uncertainty in the surface pressure since the areas of consistency in Figure 5a and 5b are within ~7% of one another. It can also be seen that our non-ideal estimates of $O_2$ fugacity are higher than those obtained if the fugacity coefficients of $H_2O$ and $H_2$ cancel out (see Equation (J3)). If we overcorrected for differing non-ideal behavior of these species via an extrapolation (Figure J1), then the oxidation state would be more reduced, supporting more N sequestration. Similarly, if the unknown magma ocean temperature were lower than Benneke et al.'s (2024) range, then the magma would be more reduced than what we stated above (Figure 5). It should also be noted that higher $H_2$ fugacities enhance N solubility (due to $NH_x$ formation; Mysen et al. 2008) probably beyond that previously calculated for sub-Neptune magma oceans (Rigby et al. 2024; Shorttle et al. 2024).

The data from TOI-270 d allow us to conclude that a magma ocean can host geochemical conditions conducive to N removal from the atmosphere. However, full self-consistency remains to be demonstrated, including the amount of melt and how it impacts the mass balance of volatile elements between atmosphere and magma reservoirs (see below). On the other hand, the present constraints on the oxidation state expressed in TOI-270 d's atmosphere as $f_{O_2}$ can be incorporated into improved models of magma ocean geochemistry on sub-Neptunes (Kite et al. 2020; Schlichting & Young 2022; Rigby et al. 2024; Shorttle et al. 2024). They can also broaden our understanding of the redox evolution of rocky planetary interiors, both near and far (Zolotov & Fegley 1999; Frost & McCammon 2008; Glein et al. 2008; Wadhwa 2008; Zolotov et al. 2013; Armstrong et al. 2019; Doyle et al. 2019; Guimond et al. 2023; Young et al. 2023).

If there is indeed a magma ocean removing nitrogen from the atmosphere of TOI-270 d, it would affect the abundances of other elements in the atmosphere. The magnitude of fractionation will depend on differences in solubilities of volatile elements in silicate melt and on the mass of melt. Although TOI-270 d is thought to have a large mass of rock relative to the mass of its atmosphere (Benneke et al. 2024), we do not know how much of the rock is molten. The apparent need to go from cosmochemically plausible bulk N/H ratios to much lower values in the atmosphere (see Table I1) implies that N is more soluble than H, and the magma ocean should be sufficiently thick to affect the mass balance of N/H. Above, we constrained the oxidation state which can feed into future efforts aimed at setting additional constraints on the properties of a magma ocean. Rigby et al. (2024) demonstrated the application of a holistic magma ocean modeling approach on K2-18 b. Given the greater amount of data on TOI-270 d from Benneke et al. (2024) and now this study, TOI-270 d would be an outstanding next target for such work (Nixon et al. 2025). A deeper understanding of fractionation on TOI-270 d may lead to a revision of our predicted enrichments in Figure 4, but we must await assessments of mass balance before further predictions can be made.

Lastly, the recognition of plausible pathways (Shorttle et al. 2024; Wogan et al. 2024; Figures 2b, 4b, and 5) that can dramatically decrease the abundance of observable $NH_3$ on temperate sub-Neptunes means that a non-detection of $NH_3$ cannot be used to discriminate between Hycean world and more conventional scenarios of a thick, hot atmosphere on these planets.

### 5. Concluding Remarks

*5.1. Summary*

This paper attempted to expand the scope of exoplanetary geochemistry, which can now be guided by compositional data from temperate sub-Neptune exoplanets (Madhusudhan et al. 2023; Benneke et al. 2024). Our overall philosophy was to develop modeling approaches that are rooted as much as possible in empirical experience. This experience includes fumaroles on Earth that constrain quench temperatures between redox species in hot gases, and making planets out of meteorites and cometary material to understand how different elements can reach different levels of enrichment in planetary atmospheres. Our approaches were simple, perhaps too simple in some cases if the goal is to accurately pinpoint the composition, present conditions, and history of the planet. If, instead, the goal is to suggest new ways of thinking about geochemistry on exoplanets that maintain focus on key variables and how they can be connected to observational data, as well as large-scale links between what we observe and how the atmosphere might have originated, then a different path to progress can be taken. The latter is the point of view we pursued.

We explored whether TOI-270 d could be an archetypal deep-atmosphere sub-Neptune (Benneke et al. 2024; Yang & Hu 2024). Adopting this premise, we found that a great deal of insight can be extracted from recent JWST data (Figure 6). Our major contributions are summarized as follows:

1) Graphical analysis combining constraints from physical and chemical models can facilitate rapid identification of the mutually consistent parameter space. We anticipate that this kind of approach will serve as a helpful initial step to narrow down the parameter space in more rigorous retrieval methods.

2) We introduced a heuristic that enables the quench temperatures of redox reactions to be estimated if the quench temperature of one reaction can be determined using observational data. On planets like TOI-270 d, where atmospheric temperatures are warm enough to prevent water condensation, $CO_2$-$CH_4$ equilibrium is well-suited for this purpose. Quenching of thermochemical equilibrium between $CH_4$, $H_2O$, $CO_2$, and $H_2$ apparently occurs at ~1-10 bar and ~900-1100 K on TOI-270 d.

3) A current inconsistency between deep-atmosphere models of temperate sub-Neptunes and observations is an expected CO abundance that is too high. We showed that this can be resolved if the atmosphere is relatively rich in $H_2O$ and poor in $CO_2$, in ways that overlap with data from TOI-270 d. Moreover, lower-temperature quenching of CO-$CO_2$ equilibrium (i.e., the water-gas shift reaction) supports a decreased CO abundance.

4) Properly assessing the atmospheric inventory of nitrogen requires quantification of $N_2$-$NH_3$ equilibrium deeper in the atmosphere (Fortney et al. 2020). We showed that $N_2$ can be the dominant form of nitrogen on a temperate sub-Neptune. From this assessment, we concluded that the atmosphere of TOI-270 d is remarkably deficit in total N.

5) Atmospheric enrichments of C, $O_{gas}$, and N (relative to protosolar H) on TOI-270 d fall on trends defined by the solar system's giant planets. This broad consistency may provide an important clue to how TOI-270 d acquired its volatile endowment.

6) We showed how the enrichments of volatile elements can be predicted using geochemical mixing models. This is an alternative to the frequently made assumption of uniform enrichments of heavy elements. Our models revealed that a rich variety of planetary building blocks mixed with primordial gas yield enrichments consistent with observed C/H and $O_{gas}$/H ratios. Accretion of LL or CO chondrites would also provide a mechanism for the planet to have a low intrinsic abundance of nitrogen.

7) We inferred upper and lower limits on the oxidation state of the near-surface atmosphere of an exoplanet from observational data for the first time (see Bello-Arufe et al. 2025). When extended to the surface, the atmospheric inventory of $O_{gas}$ implies that TOI-270 d's oxidation state is more reduced than the iron-wüstite buffer by up to several dex. Conditions were found that would impart a surface magma ocean (if present) with the thermodynamic potential to sequester a significant amount of nitrogen.

8) The inability to detect NH$_3$ on temperate sub-Neptunes has perplexed the community. The broad perspective that we tried to foster leads us to suggest that a sequence of processes may be implicated (Figure 6). First, the planet probably started out with a high C/N ratio (naturally depleted N). Then, a magma ocean at the planet's surface could remove nitrogen from the atmosphere (Shorttle et al. 2024). Subsequent speciation to N$_2$ higher in the atmosphere would remove additional N atoms from the pool of atmospheric NH$_3$. Finally, the detectability of NH$_3$ could be compromised if vertical mixing is too slow to replenish NH$_3$ destroyed by photolysis (Wogan et al. 2024).

9) Because a revised deep-atmosphere scenario can accommodate depleted CO and NH$_3$ abundances, the apparent absence of these species should no longer be taken as evidence against this type of scenario for TOI-270 d and similar planets, such as K2-18 b. Our results imply that the Hycean hypothesis is currently unnecessary to explain any data, although this does not preclude the existence of Hycean worlds.

**Figure 6.** A graphical summary of our most important findings with emphasis placed on the origin, fate, and context of "missing ammonia" in the atmosphere of TOI-270 d. Numbers indicate a sequence of processes that may result in a lack of detectable NH$_3$.

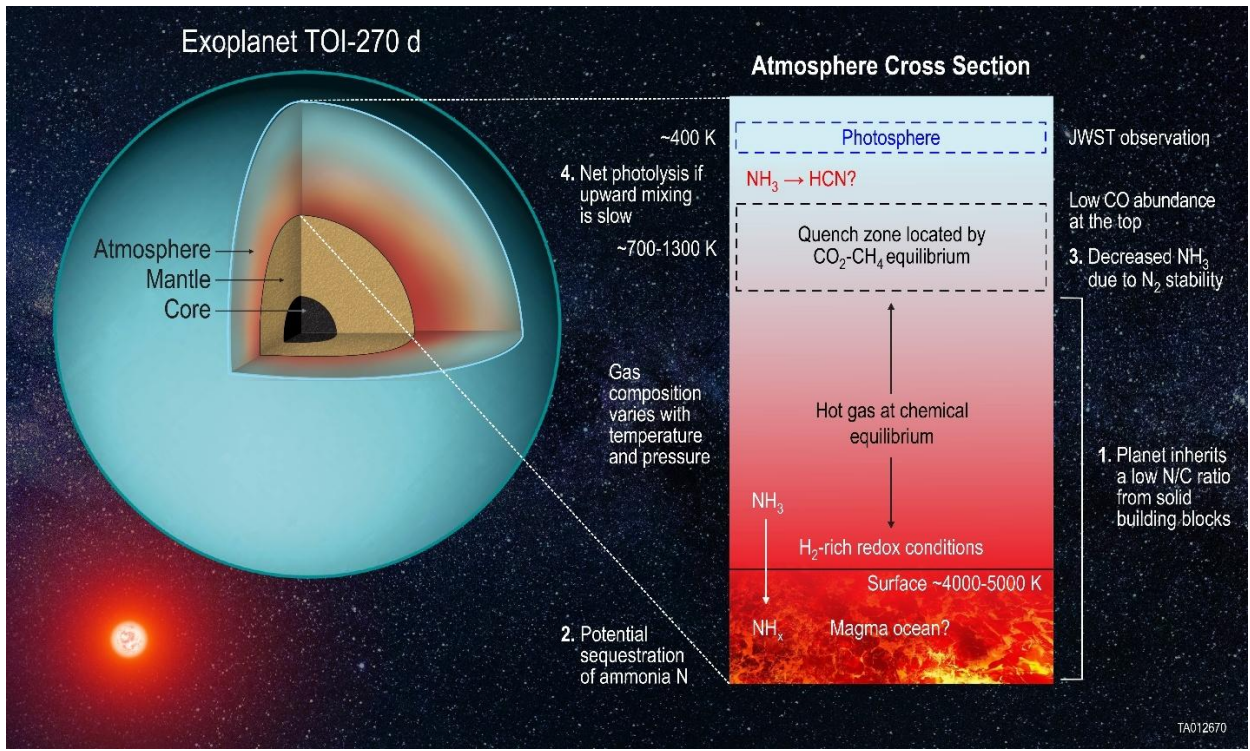

*5.2. Recommendations*

While we have learned a great deal in helping to make the present state and origin of TOI-270 d's atmosphere less enigmatic, much remains to address the limitations of this study and to build upon the frameworks introduced here. These are early days in our understanding of temperate sub-Neptunes. Here, we outline recommended next steps, grouped under four general categories, which could be taken to enable further progress:

1) *Make more observations of TOI-270 d.* This planet is a superb model for the temperate sub-Neptune population. Mass measurements have been made (Van Eylen et al. 2021; Kaye et al. 2022), allowing the bulk density to be derived. TOI-270 d also has a high transmission spectroscopy metric (124; Kempton et al. 2018; Kaye et al. 2022), facilitating compositional characterization of its atmosphere. Additional transits should be used to perform more sensitive searches for CO and $NH_3$, particularly in the mid-infrared (JWST GO 3557, PI: Nikku Madhusudhan). Our MetMiSN model can be falsified if the mole fraction of CO is <0.26%, which might indicate a Hycean world instead (Wogan et al. 2024). Establishing a stricter constraint on the $NH_3$ abundance would further constrain the atmospheric inventory of total N and allow better assessments of the sufficiency of silicate melt or liquid water as nitrogen sinks. Carriers of other elements should also be sought. HCl may be illuminating because it is extremely water soluble (e.g., Glein & Shock 2010), which could provide a new basis for distinguishing between Hycean world vs. deep-atmosphere scenarios (see Thomas & Wood 2021). Many of these observations could be made using JWST, Ariel, and the Extremely Large Telescope (ELT).

2) *Perform additional modeling for TOI-270 d*. There is a rich parameter space of possibilities remaining to be explored. These include eddy diffusion profiles and intrinsic temperatures different from those adopted by Yang & Hu (2024) (see Mukherjee et al. 2024). The effects of other processes should also be quantified including different rock-gas partitioning and escape rates of C-N-O-H-He, the potential to volatilize oxygen originally bound to Mg or Si (Kim et al. 2023; Misener et al. 2023), accretion of multiple types of solids from different disk regions, the deep, non-ideal gas speciation (which may involve the formation of H and OH radicals), and the availability of a broader range of cometary compositions (there is not just one; see Dello Russo et al. 2016). Clearly, the present paper should be seen as a starting point rather than a finished story. All of this modeling eventually needs to be made as self-consistent as possible (e.g., Krissansen-Totton et al. 2024).

    We would also like to bring attention to two more specific opportunities. First, it would be beneficial to understand what the observed S-bearing molecules (e.g., $CS_2$; Benneke et al. 2024) can tell us about the bulk abundance of sulfur (Moses et al. 2024). This would allow S to be added to our list of ingredients for making the atmosphere. This is also an interesting problem because S can partition into the metal core (if present). Second, a deeper understanding of HCN photochemistry on TOI-270 d could help to test whether slow vertical mixing is responsible for the observed $NH_3$ depletion (Benneke et al. 2024; Holmberg & Madhusudhan 2024). Results from Hu (2021) suggest that the HCN abundance should peak just above the altitude where $NH_3$ becomes strongly depleted. Thus, we might expect to see HCN in the photosphere if the nitrogen inventory is large but is no longer in $NH_3$ due to photochemistry. Holmberg & Madhusudhan (2024) retrieved a 2σ upper limit of ~3.5 ppmv on the HCN mixing ratio. Further photochemical modeling may be fruitful to link this lack of HCN to the deeper abundance of $NH_3$ as a function of the strength of mixing (e.g., $K_{zz}$).

3) *Characterize the stellar composition*. In Section 4, we normalized heavy element/hydrogen ratios with respect to the conventional protosolar ratios from Lodders (2021) (see Truong et al. 2024). Such normalization is convenient when comparing different elements. However, the protosolar assumption is only an approximation used to begin work in the absence of detailed data. A more rigorous procedure is to normalize an exoplanet's composition with respect to its host star's abundances (e.g., Welbanks et al. 2019; Sun et al. 2024). Since data on C-N-O are not available for TOI-270, we recommend that this star's photospheric abundances be measured (in the forms of CO, CN, and OH), as recently done for K2-18 using the Gemini South telescope (Hejazi et al. 2024). Gravitational settling of heavy elements from the photosphere may not be significant because TOI-270 is a low-mass star (0.39$M_\odot$; Van Eylen et al. 2021); thus, its interior should be more convective than the Sun's (Chabrier & Baraffe 1997), making its photospheric elemental ratios more representative of the star's bulk composition.

4) *Make comparisons to other planets.* TOI-270 d needs to be placed in a sufficient context to determine whether it may be representative of temperate sub-Neptunes, or if this class shows greater compositional diversity (Luque & Pallé 2022). JWST has so far shown that K2-18 b's atmosphere is compositionally similar to that of TOI-270 d (Benneke et al. 2024; Holmberg & Madhusudhan 2024), but two other sub-Neptunes look different (L98-59 d, Banerjee et al. 2024; Gressier et al. 2024; LP 791-18 c, Roy et al. 2024). Additional results should become available in the next year, which will likely include analyses of spectroscopic data from LTT 3780 c (JWST GO 3557, PI: Nikku Madhusudhan), TOI-1231 b (JWST GO 3557, PI: Nikku Madhusudhan), TOI-4336 A b (JWST GO 4711, PI: Renyu Hu), and TOI-1801 b (JWST GO 6932, PI: Rafael Luque).

Even more exciting is that NASA is planning to start a new mission to our nearest Neptune-like planet, Uranus (Mandt 2023). There is an exoplanet connection as this mission could provide insights into the chemistry of sub-Neptunes like TOI-270 d by assessing the magnitude of endogenous sources of CO and $N_2$. This opportunity will help us to test models of CO quenching in a relevant environment and provide key information related to the atmospheric enrichment of N (Fegley et al. 1991). The latter is notable as $N_2$ is very difficult to measure on sub-Neptune exoplanets (e.g., Cadieux et al. 2024), yet it can be directly measured using a mass spectrometer on an entry probe into Uranus.

**Acknowledgments**

The amazing capabilities of the James Webb Space Telescope made this work possible. C. R. G. was supported by funding from Southwest Research Institute, NASA, and the Heising-Simons Foundation. X. Y. and C. N. L. were supported by the NASA Habitable Worlds Program (Grant 80NSSC24K0075). We thank the reviewer for a number of detailed and constructive comments. We also appreciate the superb job done by Rene Limon in helping to make Figure 6. C. R. G. thanks Conel Alexander for instilling an appreciation of the importance of chondrites to understand the origin of planetary volatiles. I also wish to express my gratitude to Misha Zolotov, whose example is hopefully reflected in the quality of this work.

**Appendix A. Proposed Correlations between Quench Temperatures of Key Redox Reactions in Hot Gases that Undergo Cooling**

To understand what the implications of $CO_2$-$CH_4$ equilibrium might be for other reactions on TOI-270 d, we need to be able to convert between different quench points. As a source of empirical guidance, we consider the states of gases from fumaroles to assess the relative rates of key redox reactions. As on deep-atmosphere exoplanets, quenched ratios of fumarolic gases record processes of equilibration between chemical species occurring at depth. They can provide real-world insights into the relative equilibration rates of different reactions, serving as complementary sources of information to full-up chemical kinetic modeling. Magmatic gases can also be useful in this respect, but their higher temperatures disfavor the formation of $CH_4$ at equilibrium, making them less relevant to temperate sub-Neptunes. At White Island in New Zealand, $CO_2$-$CH_4$ equilibrium is apparently reached around 500 °C (773 K) (Giggenbach 1987). CO-$CO_2$ is more kinetically responsive and was found to equilibrate down to ~300-400 °C (573-673 K), while $N_2$-$NH_3$ is thought to have reached equilibrium at higher temperatures, perhaps as high as ~600-700 °C (873-973 K). The ordering of these systems is correct in terms of chemical reactivity, but there is some flexibility in defining specific temperatures at this stage in the analysis. Here, we nominally adopt $T_{N_2-NH_3}$ = 923 K, $T_{CO_2-CH_4}$ = 773 K, and $T_{CO-CO_2}$ = 623 K. However, the rule of thumb that we develop below using these values will be evaluated for consistency with an independent theoretical model in Appendix B.

Quench temperatures will differ in environments with different dynamical timescales. This needs to be accounted for to apply constraints from White Island to other systems containing hot gases subject to quenching. Perhaps the simplest relationship that may also be chemically reasonable is to assume that quench temperatures are proportional to one another. For equilibrium between redox species *i* and *j*, we may assume that

$$T_{i-j}(\text{K}) \propto T_{CO_2-CH_4}(\text{K}) . \tag{A1}$$

This type of relationship has the potential to show consistent behavior with temperature, although it strictly holds only if the Arrhenius pre-exponential factors are the same. Nevertheless, the functional form may be appropriate enough to yield useful approximations.

From the White Island data, we suggest the following correlations:

$$T_{CO\text{-}CO_2}(\text{K}) \approx \left(\frac{623}{773}\right) \times T_{CO_2\text{-}CH_4}(\text{K}) \approx 0.8 \times T_{CO_2\text{-}CH_4}(\text{K}) \tag{A2}$$

and

$$T_{N_2\text{-}NH_3}(\text{K}) \approx \left(\frac{923}{773}\right) \times T_{CO_2\text{-}CH_4}(\text{K}) \approx 1.2 \times T_{CO_2\text{-}CH_4}(\text{K}) . \tag{A3}$$

**Appendix B. Comparisons of Empirical Scaling of Quench Temperatures Against a More Complex Parameterization**

It would be helpful to benchmark our model of quenching (see Appendix A) against an established model. Here, we consider the model of Zahnle & Marley (2014). A word of caution: Their model was meant to be used on objects of near-solar metallicity. However, it may be informative to also perform an extrapolation and see if anything interesting emerges.

Zahnle & Marley (2014) derived an equation for the timescale (*t*) of the reaction from CO to $CH_4$. If the interconversion between $CO_2$ and CO is assumed to be significantly faster (Bustamante et al. 2004), then the timescale to reach $CO_2$-$CH_4$ equilibrium should converge to that of the CO to $CH_4$ reaction. Thus, we can write (Zahnle & Marley 2014):

$$t_{\text{CO}_2-\text{CH}_4} = \left(\frac{1}{t_{q1}} + \frac{1}{t_{q2}}\right)^{-1}, \tag{B1}$$

with

$$t_{q1}(\text{s}) = \left(1.5 \times 10^{-6}\right) P^{-1}_{\text{CO}_2-\text{CH}_4} m^{-0.7} \exp\left(42,000/T_{\text{CO}_2-\text{CH}_4}\right) \tag{B2}$$

and

$$t_{q2}(\text{s}) = (40) P^{-2}_{\text{CO}_2-\text{CH}_4} \exp\left(25,000/T_{\text{CO}_2-\text{CH}_4}\right), \tag{B3}$$

where $P$ (quench pressure) is in bars and $T$ (quench temperature) is in kelvins, $m$ stands for metallicity relative to solar, and subscripts designate the reaction being quenched. Zahnle & Marley (2014) also found analytical approximations for the quenching timescales for $CO$-$CO_2$ and $N_2$-$NH_3$:

$$t_{\text{CO}-\text{CO}_2}(\text{s}) = \left(1.0 \times 10^{-10}\right) P^{-0.5}_{\text{CO}-\text{CO}_2} \exp\left(38,000/T_{\text{CO}-\text{CO}_2}\right) \tag{B4}$$

and

$$t_{\text{N}_2-\text{NH}_3}(\text{s}) = \left(1.0 \times 10^{-7}\right) P^{-1}_{\text{N}_2-\text{NH}_3} \exp\left(52,000/T_{\text{N}_2-\text{NH}_3}\right). \tag{B5}$$

Assuming that the dynamics of vertical transport do not change significantly between the three quench points, their reaction timescales should follow

$$t_{\text{CO}_2-\text{CH}_4} \approx t_{\text{CO}-\text{CO}_2} \approx t_{\text{N}_2-\text{NH}_3}. \tag{B6}$$

A strategy for making comparisons may now be suggested. We consider the warm atmosphere profile in Figure 1 as the test subject. At each value of $T_{\text{CO}_2-\text{CH}_4}$, we compute $T_{i-j}$ for redox pair *i-j* using Equation (A2) or (A3). This is our model's quench temperature. Values of $P_{i-j}$ are estimated by interpolating the warm atmosphere curve. We calculate quench temperatures for $CO$-$CO_2$ and $N_2$-$NH_3$ based on Zahnle & Marley's (2014) model by solving Equation (B1) and setting the result equal to Equation (B4) or (B5). Metallicities of 1 (where the model was established) and 200 (more relevant to TOI-270 d; Benneke et al. 2024) times solar are explored.

Figure B1 shows how the simple model compares to an approach based on the work of Zahnle & Marley (2014). Quench temperatures from the latter model are broadly proportional to one another, as predicted. The reasonable agreement between results from our model and the more complex model suggests that appropriate values were adopted from White Island to derive scaling factors in Equations (A2) and (A3). In fact, results for another temperate sub-Neptune (K2-18 b) from a different chemical model (KINETICS) indicate $T_{\text{CO}-\text{CO}_2}/T_{\text{CO}_2-\text{CH}_4} \approx 0.76$ and $T_{\text{N}_2-\text{NH}_3}/T_{\text{CO}_2-\text{CH}_4} \approx 1.17$ (Yu et al. 2021), quite close to our values. Of course, these are only models and the agreement is not perfect, but it may be good enough. If the difference between these approaches is taken as a first-order estimate of the uncertainty of our model, then the uncertainty appears to be less than ~100 K. The overall consistency bolsters confidence that this simple model can provide representative translations between quench temperatures of key redox reactions. It also suggests that the equilibrium states of gases from fumaroles can provide helpful constraints on quenching of redox reactions in hot regions of planetary atmospheres, and rocks in the former environments do not significantly affect the relative relationships between quench temperatures, at least among these species.

**Figure B1.** How results from our approach of relating quench temperatures compare to predictions based on the parameterization of Zahnle & Marley (2014) (ZM2014) at two different metallicities. These results correspond to *P-T* conditions along the warm atmosphere profile in Figure 1. This restricts the analysis to temperatures between ~500 and ~1370 K. The solid lines were calculated independently of ZM2014 data points. The dashed lines denote where quench temperatures would be the same. (a) CO-$CO_2$ equilibrium is predicted to be quenched at lower temperatures than is $CO_2$-$CH_4$ equilibrium. (b) The opposite relationship applies to $N_2$-$NH_3$.

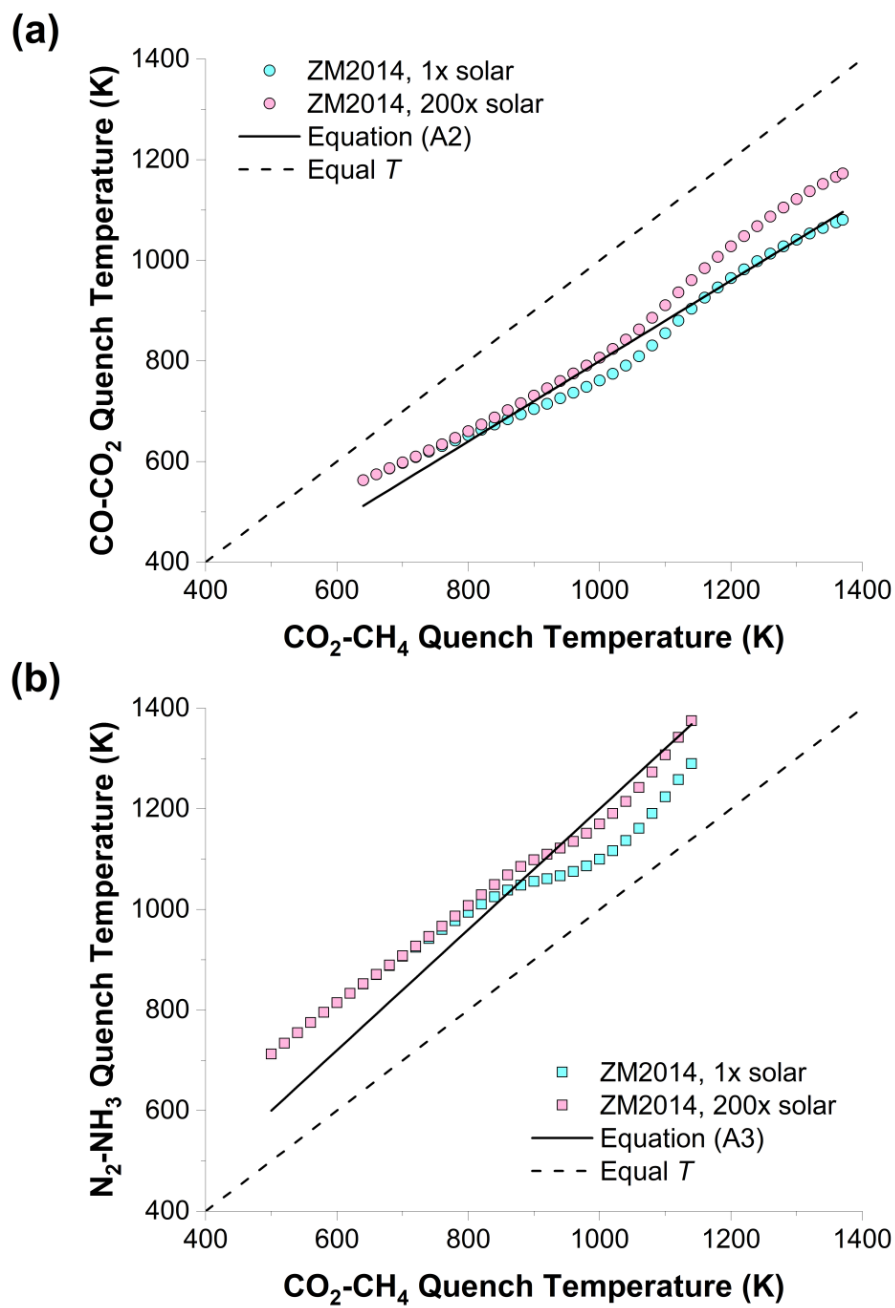

**Appendix C. Determination of Whether Conditions of $CO_2$-$CH_4$ Equilibrium are Compatible with No Detectable CO**

Carbon monoxide production in the quench region of a sub-Neptune atmosphere can be estimated by assuming chemical equilibrium according to the following reaction:

$$CO_2(g) + H_2(g) \rightleftharpoons CO(g) + H_2O(g). \tag{C1}$$

The equilibrium constant of this reaction can be expressed as

$$K_{C1} \approx \frac{y_{CO} y_{H_2O}}{y_{CO_2} y_{H_2}}. \tag{C2}$$

Because there are equal numbers of gaseous species on both sides of Reaction (C1) and no condensed phases in the reaction, the ideal-gas equilibrium state is independent of pressure. After taking the logarithm of Equation (C2) and rearranging, we have

$$\log(y_{CO}) \approx \log(K_{C1}) + \log(y_{CO_2}) + \log(y_{H_2}) - \log(y_{H_2O}). \tag{C3}$$

A regression of data from the NIST-JANAF compilation (Chase 1998) indicates that the equilibrium constant of this reaction can be represented by

$$\log(K_{C1}) = 5.57 - \frac{2319}{T(K)} - 1.137 \times \log(T(K)). \tag{C4}$$

The goodness of fit was found to be excellent with log $K$ reproduced to within ~0.01 units over the temperature range 298-1500 K.

Equation (C3) allows us to predict the mole fraction of CO in TOI-270 d's atmosphere. The predicted abundance can then be compared to the upper limit from JWST observations. The computational approach is as follows: (1) Convert the $CO_2$-$CH_4$ quench temperature to the corresponding CO-$CO_2$ quench temperature using Equation (A2); (2) use Equations (C3) and (C4), and the mole fractions listed in Table 1, to compute an initial CO mole fraction; (3) determine the mole fraction sum with CO now included (see Equation (5)); and (4) calculate a new CO mole fraction among all modeled species.

**Appendix D. Using Molecular Abundances to Set Limits on Carbon-to-Hydrogen and Oxygen-to-Hydrogen Ratios**

C/H and $O_{gas}$/H ratios in the atmosphere of TOI-270 d can be calculated using the following equations based on major species:

$$\left(\frac{C}{H}\right)_{atm} \approx \frac{y_{CH_4} + y_{CO_2} + y_{CO}}{2 y_{H_2} + 2 y_{H_2O} + 4 y_{CH_4}} \tag{D1}$$

and

$$\left(\frac{O_{gas}}{H}\right)_{atm} \approx \frac{y_{H_2O} + 2y_{CO_2} + y_{CO}}{2y_{H_2} + 2y_{H_2O} + 4y_{CH_4}}. \tag{D2}$$

We estimate limiting values of these elemental ratios in two ways. In a first approach (called "worst case"), we find the combinations of mole fractions that minimize or maximize the ratio of interest (Table D1). These combinations obey the following rules: (1) The mole fractions of $CO_2$ and $CH_4$ must fall within the 1σ ranges from JWST observations and yield a $CO_2/CH_4$ ratio consistent with the 1σ range of Benneke et al. (2024); (2) the mole fraction of $H_2O$ is allowed to vary from 0.078 to 0.16 (Table 1); (3) the $He/H_2$ ratio is given by the protosolar value (0.20; see Section 2.1); and (4) for each composition considered, there needs to be some *P-T* that allows the predicted mole fraction of CO to be below the detection limit (see Appendix C). This approach is the worst case because all of the mole fractions must conspire to push the ratio of interest in the same direction, which is unlikely. Yet, it is useful to know what this case looks like.

**Table D1.** Photospheric compositions used to define reasonable limiting values for C/H and $O_{gas}/H$ ratios in the atmosphere of TOI-270 d. See Appendix D for how we define "worst case" and "driving species."

| Quantity | C/H Lower Limit | | C/H Upper Limit | | $O_{gas}$/H Lower Limit | | $O_{gas}$/H Upper Limit | |
|---|---|---|---|---|---|---|---|---|
| | Worst Case | Driving Species | Worst Case | Driving Species | Worst Case | Driving Species | Worst Case | Driving Species |
| Mole Fraction of $H_2$ ($y_{H_2}$) | 0.68 | 0.74 | 0.59 | 0.60 | 0.74 | 0.74 | 0.60 | 0.59 |
| Mole Fraction of He ($y_{He}$) | 0.14 | 0.15 | 0.12 | 0.12 | 0.15 | 0.15 | 0.12 | 0.12 |
| Mole Fraction of $H_2O$ ($y_{H_2O}$) | 0.16 | 0.078 | 0.15 | 0.15 | 0.077 | 0.078 | 0.15 | 0.15 |
| Mole Fraction of $CH_4$ ($y_{CH_4}$) | 0.010 | 0.020 | 0.053 | 0.041 | 0.020 | 0.020 | 0.041 | 0.053 |
| Mole Fraction of $CO_2$ ($y_{CO_2}$) | 0.0054 | 0.0053 | 0.052 | 0.052 | 0.0053 | 0.0053 | 0.052 | 0.052 |
| Mole Fraction of CO ($y_{CO}$) | 0.0035 | 0.013 | 0.035 | 0.035 | 0.013 | 0.013 | 0.035 | 0.035 |
| C/H Atomic Ratio | 0.011 | 0.022 | 0.083 | 0.077 | 0.022 | 0.022 | 0.077 | 0.083 |
| C/H Geometric Mean | 0.016 | | 0.080 | | … | … | … | … |
| $O_{gas}$/H Atomic Ratio | 0.10 | 0.059 | 0.17 | 0.17 | 0.059 | 0.059 | 0.17 | 0.17 |
| $O_{gas}$/H Geometric Mean | … | … | … | … | 0.059 | | 0.17 | |

The second approach considers how a single species that exhibits substantial variation in its mole fraction affects an elemental ratio. We call it a "driving species." For TOI-270 d, $CO_2$ is always the driver as uncertainty in its abundance has the greatest impact on both the C and $O_{gas}$ budgets (Benneke et al. 2024). We vary the mole fraction of $CO_2$ over the 1σ range of Benneke et al. (2024). All other mole fractions are assigned mean values from Benneke et al. (2024), unless doing so would violate their constraint on the $CO_2/CH_4$ ratio or their upper limit on the mole fraction of CO, or if it would not permit $CO_2$-$CH_4$ equilibrium in between Yang & Hu's (2024) *P-T* profiles. In those cases, we use values that are as close to nominal as possible. Mole fractions of $H_2$ and He are estimated as done above. Table D1 shows the results of these calculations. It should be noted that some mole fractions may not perfectly match the limits reported by Benneke et al. (2024) because our values are renormalized after the unknown abundance of CO is computed.

Reasonable ranges for the ratios of C/H and $O_{gas}$/H can be defined to use for geochemical modeling. No statistical significance is being claimed here. Instead, our goal is to merely suggest starting points for further work that can be clearly traced to assumptions on the atmospheric composition. In this paper, we consider reasonable limiting values to be given by the geometric mean of values from the worst case and driving species approaches. If this practical perspective provides a sufficient approximation, then we can make useful interpretations from modeling, which will motivate more rigorous refinements of the data.

## Appendix E. Input Parameters for Nitrogen Speciation and the Effects of Intermediate Atmospheric Temperatures on the Speciation

We seek to constrain the speciation of nitrogen at conditions consistent with the observed speciation of carbon on TOI-270 d. Since the warm atmosphere profile in Figure 1 is in the solution space (and corresponds closely with the *P-T* profile in Figure 11 in Benneke et al. 2024), we decided to focus on it. Many compositions can produce the observed carbon speciation somewhere along this *P-T* profile if $y_{H_2O}$ and/or $y_{CO_2}$ are varied smoothly. Such intricacies should be studied in the future, but our main concern here is more modest – to determine limiting values of compositional parameters needed to evaluate the N speciation.

Because they occupy the edges of the solution space, the compositions shown in Table 1 can serve as suitable cases for finding these limits. However, we cannot use them as they are because they define $CO_2$-$CH_4$ quenching, and we should expect equilibrium between $N_2$ and $NH_3$ to be quenched at higher temperatures and pressures (Figure B1b). We therefore need to respeciate their bulk compositions where $N_2$-$NH_3$ quenching is predicted to occur. This will provide new mole fractions of major species, allowing us to derive values of key parameters (*A* and *B*; see Section 3.1) that control the speciation of nitrogen at appropriate *P-T* conditions.

In addition to the warm atmosphere profile, we consider an intermediate temperature profile (Figure E1), which was generated by subtracting 50 K from the warm atmosphere profile at each pressure level. This new profile is not meant to be self-consistent, but it enables exploration of the sensitivity of the nitrogen speciation to an alternative *P-T* profile occupying part of the solution space less conducive to $N_2$ production. We can test whether estimating a robust upper limit on N/H requires a smaller value than that implied by the warm atmosphere profile. Limits on the *P-T* of $CO_2$-$CH_4$ quenching are still found where the relevant temperature profile intersects the high-pressure chemical curve. They are then translated deeper into the atmosphere where $N_2$-$NH_3$ quenching can be expected to occur according to Equation (A3).

**Figure E1.** How constraints on temperatures of apparent $CO_2$-$CH_4$ and $N_2$-$NH_3$ equilibria would change if TOI-270 d's troposphere is at least 50 K cooler than the warm atmosphere profile in Figure 1. Only the high *P* case from Table 1 is shown here because pressures cannot be much lower and allow consistency with these intermediate temperatures. Predicted mole fractions of CO never exceed Benneke et al.'s (2024) detection limit along the high *P* curve.

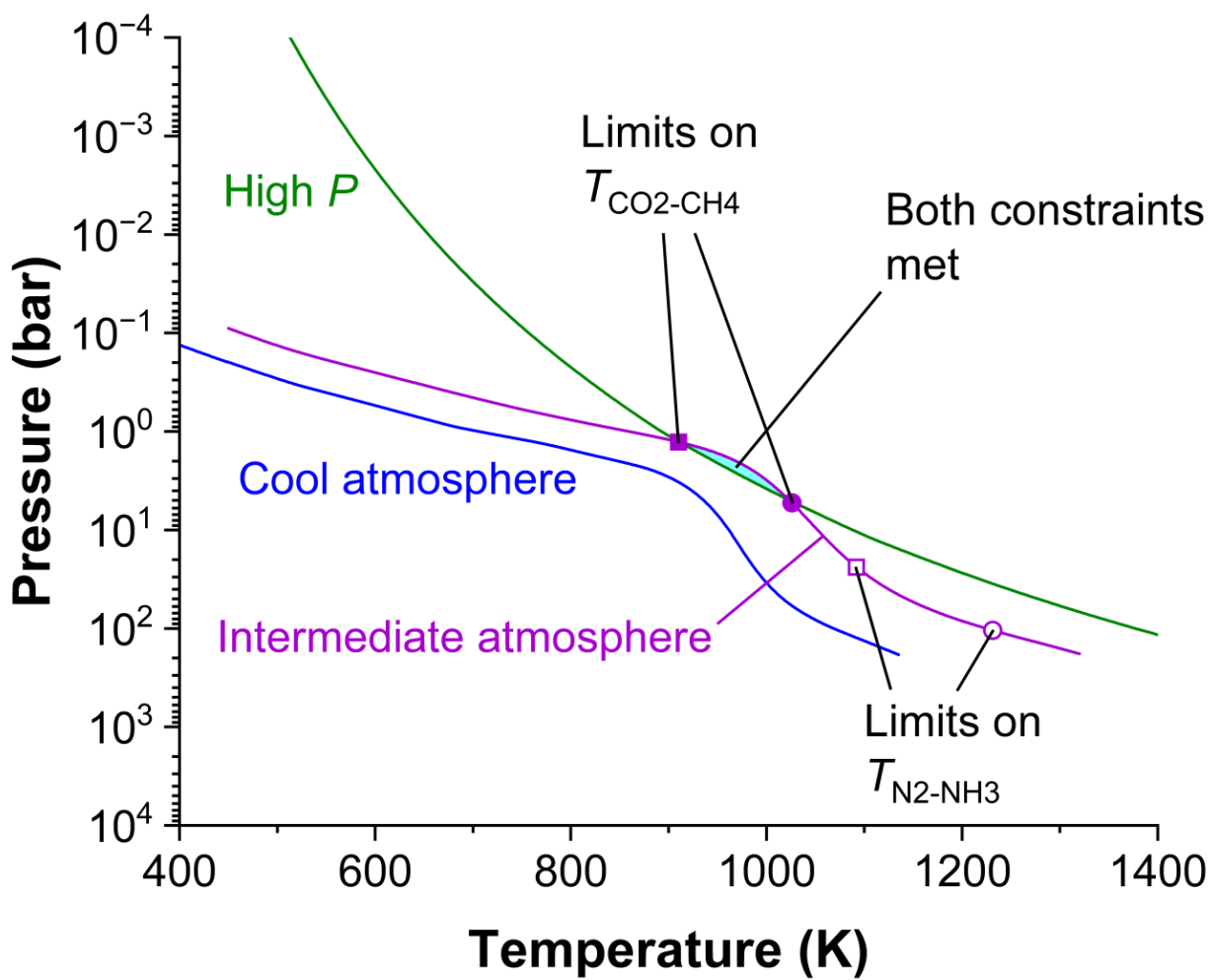

We performed calculations of ideal-gas equilibrium using the NASA Chemical Equilibrium with Applications (NASA-CEA) program (Gordon & McBride 1994; McBride & Gordon 1996). Bulk C-O-H-He compositions were derived from data in Table 1. Only the high-pressure endmember was considered in calculations for the intermediate temperature profile since the solution space is restricted to compositions near this endmember (Figure E1). The adopted temperatures and pressures were those between the open symbols along atmospheric profiles in Figures 1 and E1. The most abundant species were $H_2$, He, $H_2O$, $CH_4$, CO, and $CO_2$. The temperature dependence of $y_{H_2}$ can be seen in Figure E2a. The mole fraction of $H_2$ is the only compositional parameter that goes into calculating *A* via

$$A = \frac{K_6}{y_{H_2}^3 P^2}. \tag{E1}$$

Another key parameter is what we define as *B*. Figure E2b shows values of this parameter along the same *P-T* profiles. They were calculated based on respeciated compositions from NASA-CEA using the following equation:

$$B \approx 2y_{H_2} + 2y_{H_2O} + 4y_{CH_4}. \tag{E2}$$

**Figure E2.** Parameters needed to evaluate the nitrogen speciation (see Equations (E1) and (E2)) in the quench region of TOI-270 d. Values were computed by respeciating bulk compositions from Table 1 along our warm or intermediate temperature profile.

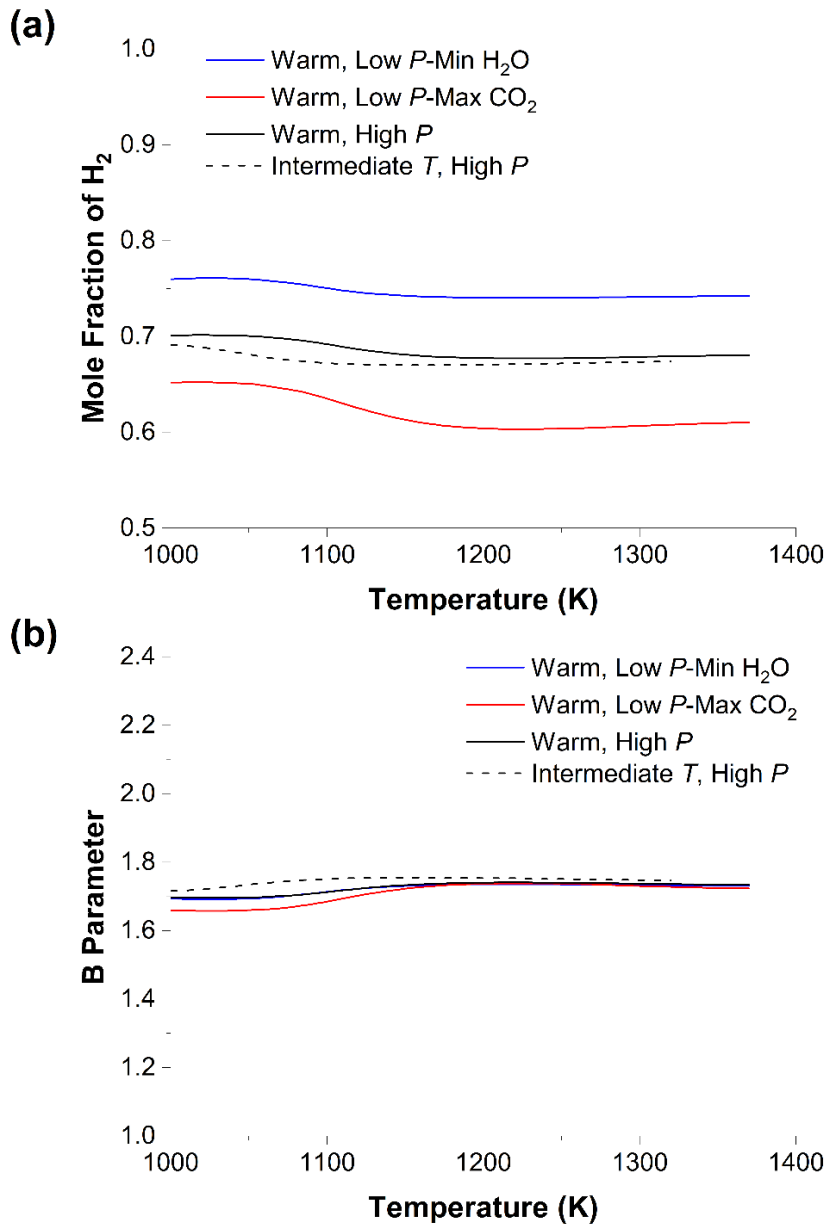

All subsequent calculations of nitrogen speciation were performed using these relationships for $y_{H_2}$ and $B$ as functions of temperature. However, it may be noticed that these parameters do not show significant variability (Figure E2), which suggests that representative values independent of temperature and composition could be adopted. As an example, we find that using $y_{H_2}$ = 0.7 and $B$ = 1.7 allows the detailed results for the mole fraction of $NH_3$ in Figure 2a to be estimated to within ~15%. One should note that these values have been vetted for TOI-270 d, and extrapolations to other sub-Neptunes are governed by *caveat emptor*.

To supplement results shown in Figure 2 for the warm atmosphere model, Figure E3 shows results along the intermediate temperature profile described above. These plots indicate how less $N_2$ production in a cooler atmosphere would decrease the upper limit on the N/H ratio required to keep the $NH_3$ abundance below the current detection limit.

**Figure E3.** (a) Abundance of $NH_3$ and (b) $N_2/NH_3$ ratio from our model of differential quenching of $N_2$-$NH_3$ and $CO_2$-$CH_4$ equilibria on TOI-270 d. These predictions are for conditions along the intermediate atmosphere profile between the open symbols in Figure E1. The C-O-H-He composition here corresponds to the high-pressure endmember in Table 1. The curve in (a) factors into how our practical upper limit on the N/H ratio is estimated (see also Figure 2a) based on the non-detection of $NH_3$ by JWST (Benneke et al. 2024).

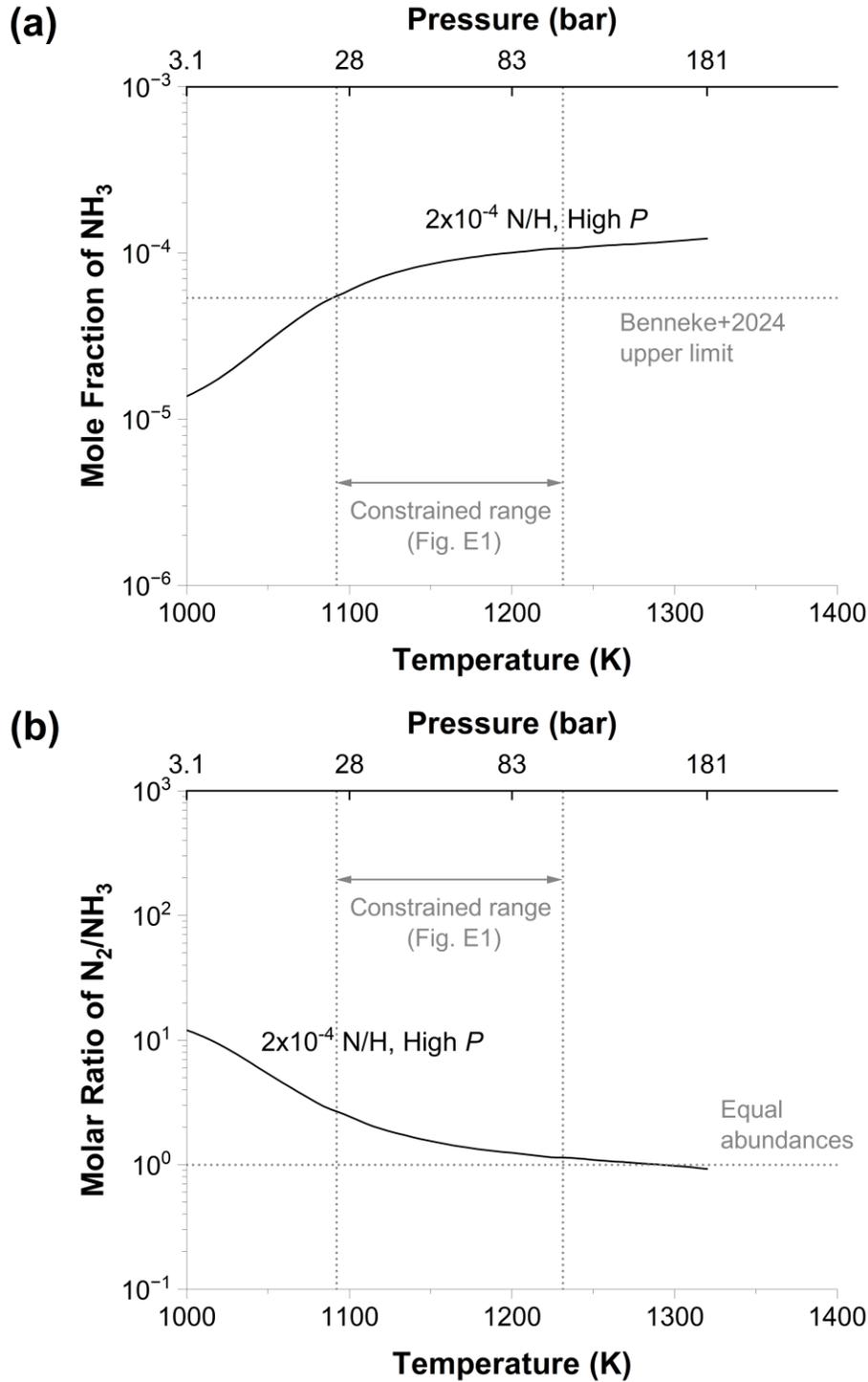

**Appendix F. Current Data on the Bulk Composition of C-H-O-N in Planetary Atmospheres**

Table F1 provides a compilation of key elemental abundance data for TOI-270 d and giant planets in the solar system.

**Table F1.** Atomic abundances of C-H-O-N in TOI-270 d's atmosphere and in the atmospheres of solar system giant planets. Values are normalized to the protosolar abundances of Lodders (2021). Italics indicate that the measurement may not have probed deep enough to avoid the effects of cloud condensation, or the value is an indirect estimate based on chemical modeling.

| Elemental Ratio | TOI-270 d | Jupiter | Saturn | Uranus | Neptune |
|---|---|---|---|---|---|
| C/H | 44-220 [a] | 2.5-4.1 [d] | 7.0-7.6 [f] | 38-62 [i] | 49-80 [l] |
| $O_{gas}$/H | 110-320 [a] | 1.4-7.7 [e] | *8.7-19* [g] | *<46* [g] *30-190* [j] | *<250* [g] *25-170* [j] |
| N/H | *<60* [a,b] *<2* [a,c] | 1.9-2.2 [e] | *1.9-3.4* [h] | 0.85-1.6 [k] | 0.65-4.8 [m] |

[a] This work.
[b] Conservative upper limit (see Section 3.2).
[c] Practical upper limit (see Section 3.2).
[d] Wong et al. (2004).
[e] Li et al. (2024).
[f] Fletcher et al. (2009).
[g] Published values (Cavalié et al. 2024) were converted from total protosolar O to $O_{gas}$.
[h] From data on upwelling of $NH_3$ at the equator (Fletcher et al. 2011) and global $NH_3$ at ~25 bar (Briggs & Sackett 1989).
[i] Sromovsky et al. (2019).
[j] Computed using the approach of Teanby et al. (2020) but converted to $O_{gas}$.
[k] Molter et al. (2021).
[l] Karkoschka & Tomasko (2011).
[m] Tollefson et al. (2021).

**Appendix G. Input Parameters for Mixing Calculations of Atmosphere-Forming Elements**

Given below are parameters that are necessary to evaluate Equations (13) and (14).

**Table G1.** Mass fractions of volatile elements in reservoirs ($w_{i,j}$) that may be analogous to source materials of sub-Neptune atmospheres. Values are mass fractions among these elements and do not include rocky elements. Planetary building blocks are shown roughly in order of increasing heliocentric distance of formation in the solar system (see Aberham & Desch 2025).

| Building Block | H | He | C | N | $O_{gas}$ |
|---|---|---|---|---|---|
| Nebular Gas | 0.710 [a] | 0.280 [a] | 3.06E-03 [a] | 8.56E-04 [a] | 5.96E-03 [a] |
| EH Chondrites | 0.0126 [b] | … | 0.101 [c] | 0.0151 [a] | 0.871 [h] |
| LL Chondrites | 0.0214 [c] | … | 0.105 [c] | 2.14E-03 [f] | 0.871 [h] |
| CV Chondrites | 0.0293 [d] | … | 0.144 [d] | 0.0166 [f] | 0.810 [h] |
| CO Chondrites | 0.0554 [d] | … | 0.145 [d] | 2.71E-03 [g] | 0.797 [h] |
| CR Chondrites | 0.0408 [d] | … | 0.0974 [d] | 6.68E-03 [f] | 0.855 [h] |
| CM Chondrites | 0.0538 [d] | … | 0.0908 [d] | 4.67E-03 [f] | 0.851 [h] |
| CI Chondrites | 0.0500 [d] | … | 0.118 [d] | 6.07E-03 [f] | 0.826 [h] |
| Comet 67P/Pluto | 0.0776 [e] | … | 0.300 [e] | 0.0148 [e] | 0.608 [e] |

Note: $O_{gas}$ represents oxygen in excess of that needed to charge balance redox-inactive metals found in silicates, notionally $Na_2O$, $K_2O$, $MgO$, $CaO$, $TiO_2$, $Al_2O_3$, and $SiO_2$.

[a] Lodders (2021).
[b] Piani et al. (2020).
[c] Alexander (2019b).
[d] Alexander (2019a).
[e] See Appendix H.
[f] Alexander et al. (2012).
[g] Alexander et al. (2018).
[h] Computed by first calculating the amount of O bound to oxide-forming metals with abundances from Alexander (2019a) or (2019b), and then subtracting that quantity from the abundance of total O listed in Lodders (2021).

**Appendix H. Estimated Bulk Composition of Cometary Material Based on Data from Comet 67P and Pluto**

A combined data set from comet 67P/Churyumov-Gerasimenko and Pluto can help to estimate a bulk composition for cometary material. Data from comets alone are insufficient because, while they provide excellent information on volatile abundances, it is challenging to relate those data to refractory components. Organic matter needs to be included to assess accurately the ratios of C/N/O/H, but the refractory/ice mass ratio in comets is currently too uncertain to be of much use (see Choukroun et al. 2020). Nevertheless, because Pluto probably formed from cometary building blocks (Glein & Waite 2018), we can use its bulk density to infer the water/rock mass ratio, which can be used to link the abundances of icy and refractory components (Truong et al. 2024).

We construct a mass balance model that relies on standard assumptions in outer planetary science. Starting with rock (= silicate + metal mixtures), we assume that its bulk composition is similar to that of CI carbonaceous chondrites (Alexander 2019a). The model rock therefore contains 12.2% Si by weight. This number is based on anhydrous rock. Proceeding with the anhydrous equivalent provides a clearer accounting of the water inventory since any phyllosilicate minerals inside Pluto can be neglected. The abundances of C-N-O-H in primordial organic matter can be calculated using molar ratios measured at comet 67P (Bardyn et al. 2017; Fray et al. 2017; Isnard et al. 2019): $C_{org}/Si \approx 5.5$, $N_{org}/C_{org} \approx 0.035$, $O_{org}/C_{org} \approx 0.3$, and $H_{org}/C_{org} \approx 1.04$.

To estimate the water/rock mass ratio, we find the proportions of water and rock that reproduce Pluto's bulk density (~1854 kg m$^{-3}$). We use a water density (~950 kg m$^{-3}$) between those of ice Ih and liquid water since Pluto appears to have an ocean (Nimmo et al. 2016). We use a density of 3500 kg m$^{-3}$ to represent anhydrous rock; this is similar to the densities of Jupiter's moon Io (~3530 kg m$^{-3}$) and asteroid Vesta (~3460 kg m$^{-3}$). Organic matter will not have a significant effect on the bulk density if its density is similar to that of Pluto. That is, the water/rock mass ratio (~0.5) will be mostly unaffected. This may be a reasonable approximation. Reynard & Sotin (2023) suggested that carbonaceous matter should have a density of ~1400-1800 kg m$^{-3}$, depending on its degree of thermal evolution. As a simplification, we neglect a minor decrease in density and assume that organic matter completely cancels out. If it does not, then we will overestimate the abundances of water and associated volatiles. The key result of this density calculation is a $H_2O/Si$ molar ratio of ~6.3.

Lastly, we account for the contributions of volatiles accreted with water. These are taken from comet 67P data (Rubin et al. 2019) to maintain consistency with our adopted values for organic matter. We consider the following relative abundances of major species in cometary ices (by mole): 100 $H_2O$: 4.7 $CO_2$: 3.1 CO: 3.1 $O_2$: 0.67 $NH_3$: 0.14 HCN: 0.089 $N_2$. With these and other values given above, we can estimate the bulk abundances of C, N, O, and H given in Table G1.

# Appendix I. Extent of Nitrogen Removal Needed to be Consistent with No Ammonia Detection by JWST

Table I1 sets requirements for the strength of a hypothesized nitrogen removal process to allow TOI-270 d to be a metal-rich, miscible-envelope sub-Neptune.

**Table I1.** Parameter space of predicted C/H and O$_{gas}$/H ratios that fall within both ranges inferred from JWST data, and depletion factors (see Equation (17)) necessary to bring corresponding N/H ratios into agreement with our upper limits (Table F1).

|  | Mutually Consistent Ranges | | Required N Depletion Factors | |
| --- | --- | --- | --- | --- |
| Cosmochemical Model | C/H with Respect to Protosolar [a] | O$_{gas}$/H with Respect to Protosolar [a] | For Conservative Upper Limit | For Practical Upper Limit |
| Equal Enrichments of Heavy Elements | 110-220 | 110-220 | >1.8 | >55 |
| EH Chondrites + Nebular Gas | 44-73 | 190-320 | No depletion needed | >12 |
| LL Chondrites + Nebular Gas | 44-76 | 180-320 | No depletion needed | >2.1 |
| CV Chondrites + Nebular Gas | 44-110 | 125-320 | No depletion needed | >9.4 |
| CO Chondrites + Nebular Gas | 44-115 | 120-320 | No depletion needed | >1.9 |
| CR Chondrites + Nebular Gas | 44-72 | 195-320 | No depletion needed | >5.7 |
| CM Chondrites + Nebular Gas | 44-67 | 210-320 | No depletion needed | >4.4 |
| CI Chondrites + Nebular Gas | 44-90 | 155-320 | No depletion needed | >4.4 |
| Comet 67P/Pluto + Nebular Gas | 105-220 | 110-230 | No depletion needed | >9.6 |

[a] In this paper, we consider protosolar C/H, O$_{gas}$/H, and N/H ratios to be $3.6 \times 10^{-4}$, $5.3 \times 10^{-4}$, and $8.7 \times 10^{-5}$, respectively (Lodders 2021).

## Appendix J. Oxidation State at the Base of TOI-270 d's Atmosphere

The oxidation state at the silicate rock/magma surface of TOI-270 d can be quantified by oxygen fugacity ($f_{O_2}$; Frost 1991). This parameter can be calculated from the molar ratio of H$_2$O/H$_2$ in that environment. The relevant reaction is:

$$2H_2O(g) \rightleftharpoons 2H_2(g) + O_2(g). \qquad (J1)$$

The equilibrium constant of this reaction can be written as

$$K_{J1} = \frac{f_{H_2}^2 f_{O_2}}{f_{H_2O}^2} = \left(\frac{\phi_{H_2}}{\phi_{H_2O}}\right)^2 \left(\frac{H_2}{H_2O}\right)^2 f_{O_2}, \qquad (J2)$$

where here $f_i$ and $\phi_i$ stand for the fugacity and fugacity coefficient of the $i$th species, respectively. The standard state of gases is defined as the ideal gas at 1 bar and any temperature. The oxygen fugacity can be calculated via

$$\log f_{O_2}(\text{bar}) = \log K_{J1} + 2\log\left(\frac{\phi_{H_2O}}{\phi_{H_2}}\right) + 2\log\left(\frac{H_2O}{H_2}\right). \tag{J3}$$

As done previously, we computed log $K$ values based on NIST-JANAF data (Chase 1998) and fit those values to a simple function, as shown below:

$$\log(K_{J1}) = 4.30 - \frac{25831}{T(\text{K})} + 0.466 \times \log(T(\text{K})). \tag{J4}$$

This equation was found to be consistent with our tabulated log $K$ values to better than ~0.02 log units from 1000 to 6000 K. Note that, by definition, log $K$ values are independent of total pressure if a reaction involves only gases.

To constrain the oxygen fugacity, estimates are required of the temperature, pressure, and ratios of fugacity coefficients and number densities at the atmosphere-silicate mantle interface. We focus on a temperature range of 4000-5000 K, as suggested by Benneke et al. (2024) for a MetMiSN scenario with a magma ocean underlying the atmosphere. The atmospheric pressure ($P_{atm}$) can be estimated using the following equation:

$$P_{atm} = \frac{m_{atm} g_{eff}}{4\pi R_{int}^2} \approx \frac{m_{atm} \kappa G M_{rock}}{4\pi R_{int}^4}, \tag{J5}$$

where $G$ = 6.6743×10$^{-11}$ N m$^2$ kg$^{-2}$, and $m_{atm} \approx 0.5 M_\oplus$ (subscript $\oplus$ represents the Earth value) and $M_{rock} \approx$ 4.3$M_\oplus$ are from Benneke et al. (2024). Since they did not provide a value for the radius of the rocky interior, we assume an Earth-like composition (consistent with a MetMiSN scenario) and use the mass-radius relation of Zeng et al. (2016) to calculate $R_{int} \approx 1.5 R_\oplus$. If the effective gravitational acceleration experienced by the whole atmosphere ($g_{eff}$) is assumed to be similar to that at the base of the atmosphere, then the correction factor $\kappa \approx 1$. In this case, $P_{atm}$ would be ~490 kbar. However, this is an overestimate because gas at higher altitudes exerts less weight.

More accurate estimates can be made by determining values of $\kappa$ based on surface pressures reported by Rigby et al. (2024) for K2-18 b. K2-18 b is a potential analogue of TOI-270 d, and these researchers developed an internal structure model for it (dedicated modeling of TOI-270 d is in progress; Nardos et al. 2025). Considering an Earth-like metal/silicate mass ratio (see above) over the range of $P$-$T$ profiles modeled by Rigby et al. (2024), we can use Equation (J5) to derive $\kappa \approx$ 0.63-0.66. Because TOI-270 d and K2-18 b are not an exact match, we adopt a wider range of 0.6-0.7. Applying these values to TOI-270 d yields $P_{atm} \approx$ 290-340 kbar (~29-34 GPa).

Determining fugacity coefficients at these pressures is a formidable challenge. Fortunately, Equation (J3) requires the ratio of H$_2$O and H$_2$ fugacity coefficients, which will partially cancel. It is also helpful that Kite et al. (2020) calculated these fugacity coefficients at magma ocean conditions, albeit only up to 30 kbar. We found that the logarithm of the ratio of their fugacity coefficients tends toward linear behavior at higher *P* (Figure J1). This suggests that it may be reasonable to extrapolate them. From 2000 to 3000 K, we obtained values of $\varphi_{H_2O}/\varphi_{H_2}$ between ~16 and ~22 at 290 kbar, and between ~26 and ~38 at 340 kbar. This ratio decreases with increasing temperature as both gases become more ideal. Therefore, above 3000 K, the ratio of $H_2O$ and $H_2$ fugacity coefficients should be smaller, but it is difficult to represent this analytically because the functional dependence is non-linear with *T*. We experimented with different values and found consistent behavior up to 5000 K with lower limits on $\varphi_{H_2O}/\varphi_{H_2}$ of ~11 and ~18 at 290 and 340 kbar, respectively. Hence, our adopted ranges of this ratio at these pressures are 11-22 and 18-38.

**Figure J1.** Ratio of $H_2O/H_2$ fugacity coefficients vs. total pressure at three temperatures relevant to sub-Neptune magma oceans. Circles show ratios computed from individual fugacity coefficients from Kite et al. (2020). Lines were fit to data between 20 and 30 kbar. Extrapolations can be performed to higher pressures using the regression equations.

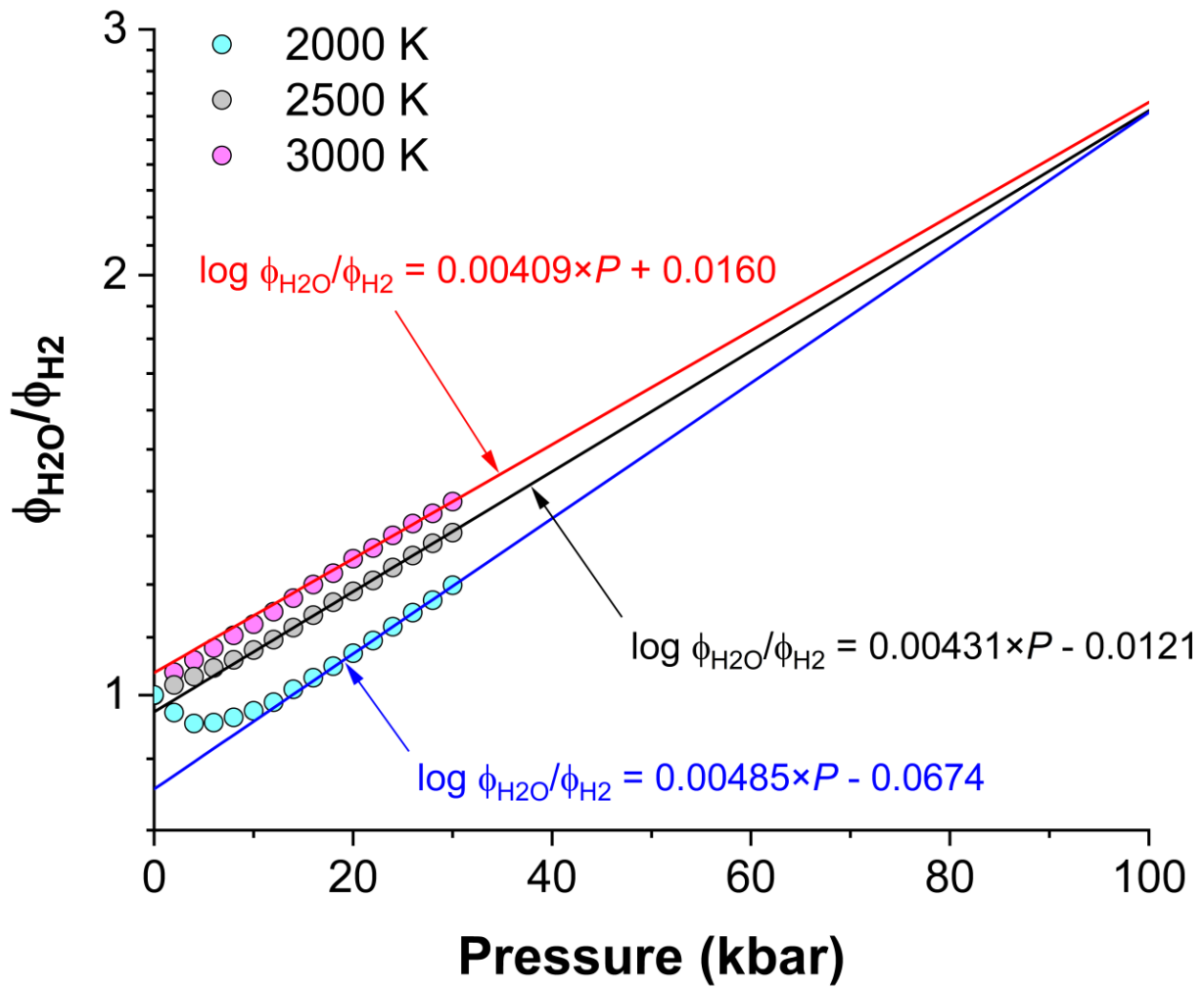

We can set limits on $f_{O_2}$ at the planet's surface by considering cases with a maximum or minimum $H_2O/H_2$ ratio. If TOI-270 d has a miscible envelope, then the bulk gas composition at the surface should be the same as that higher up (Benneke et al. 2024). However, the chemical speciation will be different since the *P-T* conditions are different. Nevertheless, broad constraints can be elucidated based on mass balance arguments. For the upper-limit case, $O_{gas}/H \approx 0.17$ (Table D1). An oxidized endmember corresponds to all $O_{gas}$ in $H_2O$ and all C in $CH_4$ (to maximize $H_2$ removal). This allows us to simplify Equation (D2), resulting in

$$\left(\frac{H_2O}{H_2}\right)_{surf} \leq \frac{\left(\frac{O_{gas}}{H}\right)_{atm}}{0.5 - \left(\frac{O_{gas}}{H}\right)_{atm} \times \left(1 + 2\left(\frac{C}{O}\right)_{atm}\right)}, \quad (J6)$$

where the upper limit on the atmospheric ratio of C/O is ~0.48 (Table D1). We can then use Equation (J6) to calculate log $(H_2O/H_2)_{surf} \lesssim 0$.

Conversely, a lower limit on $f_{O_2}$ can be computed from our lower limit on the $O_{gas}/H$ ratio of 0.059 (Table D1) by assuming that all C is in CO. We favor CO over $CO_2$ as the form of carbon that could remove oxygen because high temperatures in the deep atmosphere can be expected to produce high $CO/CO_2$ ratios (see Equation (C4)); to first order, increased pressure is unable to counteract this effect (see Equation (C2)). By combining Equations (D1) and (D2), the $H_2O/H_2$ ratio can be derived for this reduced endmember:

$$\left(\frac{H_2O}{H_2}\right)_{surf} \geq \frac{\left(\frac{O_{gas}}{H}\right)_{atm} - \left(\frac{C}{H}\right)_{atm}}{0.5 + \left(\frac{C}{H}\right)_{atm} - \left(\frac{O_{gas}}{H}\right)_{atm}}. \quad (J7)$$

We find log $(H_2O/H_2)_{surf} \gtrsim -1.1$ for this case with $(C/H)_{atm} = 0.022$ (Table D1).

Limiting values of log $f_{O_2}$ at the surface of TOI-270 d can be calculated using Equations (J3) and (J4) with the parameter values given above.

**References**

Aberham, V., & Desch, S. J. 2025, LPSC Meeting, 56, 2419
Alexander, C. M. O'D. 2019a, GeCoA, 254, 277
Alexander, C. M. O'D. 2019b, GeCoA, 254, 246
Alexander, C. M. O'D., Bowden, R., Fogel, M. L., et al. 2012, Sci, 337, 721
Alexander, C. M. O'D., Greenwood, R. C., Bowden, R., et al. 2018, GeCoA, 221, 406
Allègre, C. J., Staudacher, T., & Sarda, P. 1987, E&PSL, 81, 127
Armstrong, K., Frost, D. J., McCammon, C. A., et al. 2019, Sci, 365, 903
Atreya, S. K., Hofstadter, M. D., Reh, K. R., et al. 2019, AcAau, 162, 266
Banerjee, A., Barstow, J. K., Gressier, A., et al. 2024, ApJL, 975, L11
Bardyn, A., Baklouti, D., Cottin, H., et al. 2017, MNRAS, 469, S712
Bean, J. L., Raymond, S. N., & Owen, J. E. 2021, JGRE, 126, e2020JE006639
Bello-Arufe, A., Damiano, M., Bennett, K. A., et al. 2025, ApJL, 980, L26
Benneke, B., Roy, P.-A., Coulombe, L.-P., et al. 2024, arXiv, 2403.03325v1
Bernadou, F., Gaillard, F., Füri, E., et al. 2021, ChGeo, 573, 120192
Bitsch, B., Raymond, S. N., Buchhave, L. A., et al. 2021, A&A, 649, L5
Briggs, F. H., & Sackett, P. D. 1989, Icar, 80, 77
Bustamante, F., Enick, R. M., Cugini, A. V., et al. 2004, AIChE, 50, 1028
Cabot, S. H. C., Madhusudhan, N., Constantinou, S., et al. 2024, ApJL, 966, L10
Cadieux, C., Plotnykov, M., Doyon, R., et al. 2024, ApJL, 960, L3

Cavalié, T., Lunine, J., Mousis, O., et al. 2024, SSRv, 220, 8
Chabrier, G., & Baraffe, I. 1997, A&A, 327, 1039
Charnay, B., Blain, D., Bézard, B., et al. 2021, A&A, 646, A171
Chase, M. W., Jr. 1998, NIST-JANAF Thermochemical Tables, Parts I and II (4th ed.; Woodbury, NY: American Institute of Physics)
Choukroun, M., Altwegg, K., Kührt, E., et al. 2020, SSRv, 216, 44
Cooke, G. J., & Madhusudhan, N. 2024, ApJ, 977, 209
Dauphas, N. 2017, Natur, 541, 521
Dello Russo, N., Kawakita, H., Vervack, R. J., et al. 2016, Icar, 278, 301
Doyle, A. E., Young, E. D., Klein, B., et al. 2019, Sci, 366, 356
Elkins-Tanton, L. T., & Seager, S. 2008, ApJ, 685, 1237
Fegley, B., Gautier, D., Owen, T., et al. 1991, in Uranus, ed. J. T. Bergstralh et al. (Tucson, AZ: The University of Arizona Press), 147
Fegley, B., Zolotov, M. Y., & Lodders, K. 1997, Icar, 125, 416
Fletcher, L. N., Orton, G. S., Teanby, N. A., et al. 2009, Icar, 199, 351
Fletcher, L. N., Baines, K. H., Momary, T. W., et al. 2011, Icar, 214, 510
Fonte, S., Turrini, D., Pacetti, E., et al. 2023, MNRAS, 520, 4683
Fortney, J. J., Mordasini, C., Nettelmann, N., et al. 2013, ApJ, 775, 80
Fortney, J. J., Visscher, C., Marley, M. S., et al. 2020, AJ, 160, 288
Fray, N., Bardyn, A., Cottin, H., et al. 2017, MNRAS, 469, S506
Fressin, F., Torres, G., Charbonneau, D., et al. 2013, ApJ, 766, 81
Frost, B. R. 1991, RvMG, 25, 1
Frost, D. J., & McCammon, C. A. 2008, AREPS, 36, 389
Giggenbach, W. F. 1987, ApGC, 2, 143
Glein, C. R. 2024, ApJL, 964, L19
Glein, C. R., & Shock, E. L. 2010, GeoRL, 37, L09204
Glein, C. R., & Waite, J. H. 2018, Icar, 313, 79
Glein, C. R., Zolotov, M. Y., & Shock, E. L. 2008, Icar, 197, 157
Glein, C. R., Grundy, W. M., Lunine, J. I., et al. 2024, Icar, 412, 115999
Gordon, S., & McBride, B. J. 1994, Computer Program for Calculation of Complex Chemical Equilibrium Compositions and Applications: I. Analysis (NASA Ref. Publ. 1311), https://ntrs.nasa.gov/citations/19950013764
Gressier, A., Espinoza, N., Allen, N. H., et al. 2024, ApJL, 975, L10
Guillot, T., & Hueso, R. 2006, MNRAS, 367, L47
Guillot, T., Fletcher, L. N., Helled, R., et al. 2023, in Protostars and Planets VII, ed. S. Inutsuka et al. (San Francisco, CA: Astronomical Society of the Pacific), 947
Guimond, C. M., Shorttle, O., Jordan, S., et al. 2023, MNRAS, 525, 3703
Günther, M. N., Pozuelos, F. J., Dittmann, J. A., et al. 2019, NatAs, 3, 1099
Hejazi, N., Crossfield, I. J. M., Souto, D., et al. 2024, ApJ, 973, 31
Hirschmann, M. M. 2021, GeCoA, 313, 74
Holmberg, M., & Madhusudhan, N. 2024, A&A, 683, L2
Hopp, T., Dauphas, N., Abe, Y., et al. 2022, SciA, 8, eadd8141
Hu, R. 2021, ApJ, 921, 27
Huang, Z., Yu, X., Tsai, S.-M., et al. 2024, ApJ, 975, 146
Huber, M. L., Lemmon, E. W., Bell, I. H., et al. 2022, Ind. Eng. Chem. Res., 61, 15,449
Innes, H., Tsai, S.-M., & Pierrehumbert, R. T. 2023, ApJ, 953, 168
Irwin, P. G. J., Toledo, D., Garland, R., et al. 2018, NatAs, 2, 420
Irwin, P. G. J., Toledo, D., Garland, R., et al. 2019, Icar, 321, 550

Isnard, R., Bardyn, A., Fray, N., et al. 2019, A&A, 630, A27
Kane, S. R., Arney, G. N., Byrne, P. K., et al. 2021, JGRE, 126, e2020JE006643
Karkoschka, E., & Tomasko, M. G. 2011, Icar, 211, 780
Kasting, J. F. 1982, JGRC, 87, 3091
Kaye, L., Vissapragada, S., Günther, M. N., et al. 2022, MNRAS, 510, 5464
Kempton, E. M.-R., Bean, J. L., Louie, D. R., et al. 2018, PASP, 130, 114401
Kim, T., Wei, X., Chariton, S., et al. 2023, PNAS, 120, e2309786120
Kite, E. S., & Schaefer, L. 2021, ApJL, 909, L22
Kite, E. S., Fegley, B., Schaefer, L., et al. 2020, ApJ, 891, 111
Komacek, T. D., Showman, A. P., & Parmentier, V. 2019, ApJ, 881, 152
Kreidberg, L., Bean, J. L., Désert, J.-M., et al. 2014, ApJL, 793, L27
Krissansen-Totton, J., Wogan, N., Thompson, M., et al. 2024, NatCo, 15, 8374
Leconte, J., Spiga, A., Clément, N., et al. 2024, A&A, 686, A131
Li, C., Allison, M., Atreya, S., et al. 2024, Icar, 414, 116028
Lodders, K. 2004, ApJ, 611, 587
Lodders, K. 2021, SSRv, 217, 44
Lucazeau, F. 2019, GGG, 20, 4001
Luo, H., Dorn, C., & Deng, J. 2024, NatAs, 8, 1399
Luque, R., & Pallé, E. 2022, Sci, 377, 1211
Luu, C. N., Yu, X., Glein, C. R., et al. 2024, ApJL, 977, L51
Madhusudhan, N., Amin, M. A., & Kennedy, G. M. 2014, ApJL, 794, L12
Madhusudhan, N., Piette, A. A. A., & Constantinou, S. 2021, ApJ, 918, 1
Madhusudhan, N., Sarkar, S., Constantinou, S., et al. 2023, ApJL, 956, L13
Malamud, U., Podolak, M., Podolak, J. I., et al. 2024, Icar, 421, 116217
Malsky, I., Rogers, L., Kempton, E. M.-R., et al. 2023, NatAs, 7, 57
Mandt, K. E. 2023, Sci, 379, 640
Marty, B. 2012, E&PSL, 313-314, 56
McBride, B. J., & Gordon, S. 1996, Computer Program for Calculation of Complex Chemical Equilibrium Compositions and Applications: II. Users Manual and Program Description (NASA Ref. Publ. 1311), https://ntrs.nasa.gov/citations/19960044559
Mikal-Evans, T., Madhusudhan, N., Dittmann, J., et al. 2023, AJ, 165, 84
Miller, K. E., Glein, C. R., & Waite, J. H. 2019, ApJ, 871, 59
Misener, W., Schlichting, H. E., & Young, E. D. 2023, MNRAS, 524, 981
Mollière, P., Molyarova, T., Bitsch, B., et al. 2022, ApJ, 934, 74
Molter, E. M., de Pater, I., Luszcz-Cook, S., et al. 2021, PSJ, 2, 3
Mordasini, C., van Boekel, R., Mollière, P., et al. 2016, ApJ, 832, 41
Moses, J. I. 2014, RSPTA, 372, 20130073
Moses, J. I., Visscher, C., Fortney, J. J., et al. 2011, ApJ, 737, 15
Moses, J. I., Line, M. R., Visscher, C., et al. 2013, ApJ, 777, 34
Moses, J., Tsai, S.-M., Fortney, J., et al. 2024, DPS Meeting, 56, 308.06
Mousis, O., Cavalié, T., Lunine, J. I., et al. 2024, SSRv, 220, 44
Mukherjee, S., Fortney, J. J., Wogan, N. F., et al. 2024, arXiv, 2410.17169v1
Müller, S., & Helled, R. 2024, ApJ, 967, 7
Mysen, B. O., Yamashita, S., & Chertkova, N. 2008, AmMin, 93, 1760
Nardos, B., Nixon, M., Kempton, E., et al. 2025, AAS Meeting, 245, 314.02
Nimmo, F., Hamilton, D. P., McKinnon, W. B., et al. 2016, Natur, 540, 94
Nixon, M., Ih, J., Kempton, E., et al. 2025, AAS Meeting, 245, 314.05
Öberg, K. I., Murray-Clay, R., & Bergin, E. A. 2011, ApJL, 743, L16

Ohno, K., & Fortney, J. J. 2023a, ApJ, 946, 18
Ohno, K., & Fortney, J. J. 2023b, ApJ, 956, 125
Owen, T., & Encrenaz, T. 2006, P&SS, 54, 1188
Owen, T., Mahaffy, P., Niemann, H. B., et al. 1999, Natur, 402, 269
Piani, L., Marrocchi, Y., Rigaudier, T., et al. 2020, Sci, 369, 1110
Pierrehumbert, R., & Gaidos, E. 2011, ApJL, 734, L13
Pollack, J. B., Hubickyj, O., Bodenheimer, P., et al. 1996, Icar, 124, 62
Prinn, R. G., & Barshay, S. S. 1977, Sci, 198, 1031
Reynard, B., & Sotin, C. 2023, E&PSL, 612, 118172
Rigby, F. E., Pica-Ciamarra, L., Holmberg, M., et al. 2024, ApJ, 975, 101
Rogers, L. A., & Seager, S. 2010, ApJ, 716, 1208
Roy, P.-A., Benneke, B., Piaulet, C., et al. 2024, ExSS Meeting, 5, 502.04
Rubin, M., Altwegg, K., Balsiger, H., et al. 2019, MNRAS, 489, 594
Rustamkulov, Z., Sing, D. K., Mukherjee, S., et al. 2023, Natur, 614, 659
Schaefer, L., & Fegley, B. 2010, Icar, 208, 438
Schlichting, H. E., & Young, E. D. 2022, PSJ, 3, 127
Schmidt, S. P., MacDonald, R. J., Tsai, S.-M., et al. 2025, arXiv, 2501.18477v1
Seewald, J. S., Zolotov, M. Y., & McCollom, T. 2006, GeCoA, 70, 446
Sekine, Y., Sugita, S., Shido, T., et al. 2006, M&PS, 41, 715
Shorttle, O., Jordan, S., Nicholls, H., et al. 2024, ApJL, 962, L8
Sromovsky, L. A., Karkoschka, E., Fry, P. M., et al. 2019, Icar, 317, 266
Suer, T.-A., Jackson, C., Grewal, D. S., et al. 2023, FrEaS, 11, 1159412
Sun, Q., Wang, S. X., Welbanks, L., et al. 2024, AJ, 167, 167
Swain, M. R., Hasegawa, Y., Thorngren, D. P., et al. 2024, SSRv, 220, 61
Teanby, N. A., Irwin, P. G. J., Moses, J. I., et al. 2020, RSPTA, 378, 20190489
Thomas, R. W., & Wood, B. J. 2021, GeCoA, 294, 28
Thompson, M. A., Telus, M., Schaefer, L., et al. 2021, NatAs, 5, 575
Tollefson, J., de Pater, I., Molter, E. M., et al. 2021, PSJ, 2, 105
Truong, N., Glein, C. R., & Lunine, J. I. 2024, ApJ, 976, 14
Tsai, S.-M., Lyons, J. R., Grosheintz, L., et al. 2017, ApJS, 228, 20
Tsai, S.-M., Kitzmann, D., Lyons, J. R., et al. 2018, ApJ, 862, 31
Tsai, S.-M., Innes, H., Lichtenberg, T., et al. 2021, ApJL, 922, L27
Van Eylen, V., Astudillo-Defru, N., Bonfils, X., et al. 2021, MNRAS, 507, 2154
Visscher, C., & Moses, J. I. 2011, ApJ, 738, 72
Wadhwa, M. 2008, RvMG, 68, 493
Weidenschilling, S. J., & Lewis, J. S. 1973, Icar, 20, 465
Welbanks, L., Madhusudhan, N., Allard, N. F., et al. 2019, ApJL, 887, L20
Wogan, N. F., Batalha, N. E., Zahnle, K. J., et al. 2024, ApJL, 963, L7
Wong, M. H., Mahaffy, P. R., Atreya, S. K., et al. 2004, Icar, 171, 153
Yang, J., & Hu, R. 2024, ApJL, 971, L48
Young, E. D., Shahar, A., & Schlichting, H. E. 2023, Natur, 616, 306
Yu, X., Moses, J. I., Fortney, J. J., et al. 2021, ApJ, 914, 38
Zahnle, K. J., & Marley, M. S. 2014, ApJ, 797, 41
Zeng, L., Sasselov, D. D., & Jacobsen, S. B. 2016, ApJ, 819, 127
Zolotov, M. Y., & Fegley, B. 1999, Icar, 141, 40
Zolotov, M. Y., & Fegley, B. 2000, GRL, 27, 2789
Zolotov, M. Y., Sprague, A. L., Hauck, S. A., et al. 2013, JGRE, 118, 138